\documentclass[useAMS,usenatbib]{mn2e}
\usepackage{graphicx,epsfig}
\usepackage{multirow}
\usepackage{amsmath}
\usepackage{amssymb}
\usepackage[graphicx]{realboxes}
\usepackage{color}
\usepackage{float}
\usepackage[colorlinks=true,citecolor=blue]{hyperref}
\RequirePackage{rotating}
\def\aaps{\ref@jnl{A\&AS}}
\def\aap{\ref@jnl{A\&A}}
\def\apjl{\ref@jnl{ApJ}}
\def\apj{\ref@jnl{ApJ}}
\def\apjs{\ref@jnl{ApJS}}
\def\aj{\ref@jnl{AJ}}

\title[Photometric-z in Massive Clusters with HST.]{CLASH: Accurate Photometric Redshifts with 14 HST bands in Massive Galaxy Cluster Cores.}
\author[Molino et al.]{A. Molino$^{1,2}$, N. Ben\'itez$^2$, B. Ascaso$^3$, D. Coe$^4$, M. Postman$^4$, S. Jouvel$^5$, O. Host$^6$,   
\newauthor{O. Lahav$^5$, S. Seitz$^{7,8}$, E. Medezinski$^{9}$, P. Rosati$^{10}$, W. Schoenell$^{1,2}$, A. Koekemoer$^4$, } 
\newauthor{Y. Jimenez-Teja$^{11}$, T. Broadhurst$^{12,13}$, P. Melchior$^{9}$, I. Balestra$^{7}$, M. Bartelmann$^{14}$, }
\newauthor{R. Bouwens$^{15}$, L. Bradley$^4$,  N. Czakon$^{16}$, M. Donahue$^{17}$, H. Ford$^{18}$, O. Graur$^{19}$, } 
\newauthor{G. Graves$^{20}$, C. Grillo$^{21}$, L. Infante$^{22}$, S. W. Jha$^{23}$, D. Kelson$^{24}$, R. Lazkoz$^{14}$, } 
\newauthor{D. Lemze$^{18}$, D. Maoz$^{25}$, A. Mercurio$^{26}$, M. Meneghetti$^{27}$, J. Merten$^{28}$, L. Moustakas$^{29}$, } 
\newauthor{M. Nonino$^{30}$, S. Orgaz$^{18}$,  A. Riess$^{18}$, S. Rodney$^{31}$, J. Sayers$^{32}$, K. Umetsu$^{33}$, }
\newauthor{W. Zheng$^{18}$, A. Zitrin$^{34}$}\\ \\ Affiliations can be found after the references.}

\begin{document}
\label{firstpage}
\maketitle{}
\begin{abstract}
We present accurate photometric redshifts for galaxies observed by the Cluster Lensing and Supernova survey with Hubble (CLASH). CLASH observed 25 massive galaxy cluster cores with the Hubble Space Telescope in 16 filters spanning 0.2 - 1.7 $\mu$m. Photometry in such crowded fields is challenging. Compared to our previously released catalogs, we make several improvements to the photometry, including smaller apertures, ICL subtraction, PSF matching, and empirically measured uncertainties. We further improve the Bayesian Photometric Redshift (BPZ) estimates by adding a redder elliptical template and by inflating the photometric uncertainties of the brightest galaxies. The resulting photometric redshift accuracies are dz/(1+z) $\sim$ 0.8\%, 1.0\%, and 2.0\% for galaxies with I-band F814W AB magnitudes $<$ 18, 20, and 23, respectively. These results are consistent with our expectations. They improve on our previously reported accuracies by a factor of 4 at the bright end and a factor of 2 at the faint end.  Our new catalog includes 1257 spectroscopic redshifts, including 382 confirmed cluster members. We also provide stellar mass estimates. Finally, we include lensing magnification estimates of background galaxies based on our public lens models. Our new catalog of all 25 CLASH clusters is available via MAST. The analysis techniques developed here will be useful in other surveys of crowded fields, including the Frontier Fields and surveys carried out with J-PAS and JWST.
\end{abstract}


\section{Introduction}
\label{intro} 

The Cluster Lensing And Supernovae survey with Hubble (CLASH\footnote{\textcolor{blue}{https://archive.stsci.edu/prepds/clash/}}, \cite{2012ApJS..199...25P}, hereafter P12) is a Multi-Cycle Treasury programme awarded with 524 $HST$ orbits to image the cores of 25 massive galaxy clusters at intermediate redshifts (0.1$<$z$<$0.9). The cluster selection includes 20 X-ray selected dynamically-relaxed systems plus 5 additional specifically-selected strong lensing clusters. CLASH has combined the high spatial-resolution imaging from Hubble Space Telescope (HST) with a 16-band filter system optimized for photometric redshift estimations (4 WFC3/UVIS + 5 WFC3/IR + 7 ACS/WFC) and a typical photometric depth of 20 orbits per cluster. The combination of these three elements has made the CLASH survey an unprecedented legacy dataset.

\vspace{0.2cm}

Starting in 2010, CLASH has successfully achieved most of its main science goals: 1. Measuring the profiles and substructures of Dark Matter (DM) in galaxy clusters with unprecedented precision and resolution (\citealt{2011ApJ...742..117Z}; \citealt{2012ApJ...755...56U}; \citealt{2012ApJ...757...22C}; \citealt{2012ApJ...749...97Z}; \citealt{2012ApJ...752..141L}; \citealt{2013ApJ...769...13U}; \citealt{2013ApJ...777...43M}; \citealt{2013ApJ...762L..30Z}; \citealt{2013ApJ...774..124E}; \citealt{2014ApJ...795..163U}; \citealt{2014ApJ...786...11G}; \citealt{2014ApJ...797...34M}; \citealt{2015ApJ...806....4M}; \citealt{2015ApJ...801...44Z}; \citealt{2016ApJ...821..116U}; among others), 2. Detecting and characterizing some of the most distant galaxies yet discovered at z$>$7 (\citealt{2012Natur.489..406Z}; \citealt{2012ApJ...747....3B}; \citealt{2012ApJ...747L...9Z}; \citealt{2013ApJ...762...32C}; \citealt{2013A&A...559L...9B}; \citealt{2014ApJ...795..126B}; \citealt{2014MNRAS.438.1417M}; \citealt{2014ApJ...792...76B}; \citealt{2015ApJ...804...11P}), 3. Detecting Type-Ia supernovae (SNe-Ia) out to redshift z$\sim$2.5 to measure the time dependence of the dark energy equation of state and potential evolutionary effects in the SNe themselves (\citealt{2012ApJ...746....5R}; \citealt{2013ApJ...768..166J}; \citealt{2014ApJ...783...28G}; \citealt{2014ApJ...786....9P}; \citealt{2015ApJ...813...93S}; \citealt{2016ApJ...820...50R}) and 4. Studying the internal structure and evolution of the galaxies in and behind these clusters (\citealt{2012ApJS..199...25P}; \citealt{2013A&A...558A...1B}; \citealt{2014A&A...571A..80A}; \citealt{2014ApJ...783L..11S}; \citealt{2014A&A...565A.126P}; \citealt{2015ApJ...800...38G}; \citealt{2015ApJ...805..177D}; \citealt{2015A&A...579A...4G}; \citealt{2015ApJ...813..117F}; \citealt{2016A&A...587A..80C}; \citealt{2016A&A...585A.160A}; \citealt{2016ApJ...819...36D}; \citealt{2016ApJS..224...33B}; \citealt{2016A&A...590A.108M}; \citealt{2016JCAP...04..023P}) in combination with a spectroscopic follow-up provided by the CLASH-VLT Large Programme \citep{2014Msngr.158...48R} and wide-field deep multi-band (ugriz) ground-based Subaru/Suprime-Cam imaging (\citealt{2012ApJ...755...56U}, \citealt{2014ApJ...795..163U}). 

\vspace{0.2cm}

However, questions remain regarding the photometric redshifts: 1. to understand the unexpected underperformance of the CLASH photo-z and 2. the acquisition of a complete and reliable photo-z catalogue for cluster galaxies in the CLASH fields. As explained in P12, based on simulations of the CLASH filters and exposure times, the photo-z performance was expected to be $\delta$z$\sim$0.02(1+$z_{s}$) for 80\% of objects with magnitudes F775W$<$26 AB. Although comparable results have been achieved by similar multi-band photometric surveys (ALHAMBRA; \citealt{2008AJ....136.1325M}; \citealt{2014MNRAS.441.2891M}), the predictions stated in P12 were in disagreement with the results presented in \cite{2014A&A...562A..86J} (hereafter J14) by almost a factor of 2 (i.e., $\delta$z$\sim$0.04(1+$z_{s}$)). As emphasized in that paper, although not as precise as originally expected, a $\sim$4\% precision for the CLASH photo-z may have a subdominant effect on the mass modeling when compared to the uncertainties associated with lensing by large-scale structure along the line of sight, supporting the reliability of conducting such analysis. 

\vspace{0.2cm}

\begin{figure}
\begin{center}
\includegraphics[width=8.0cm]{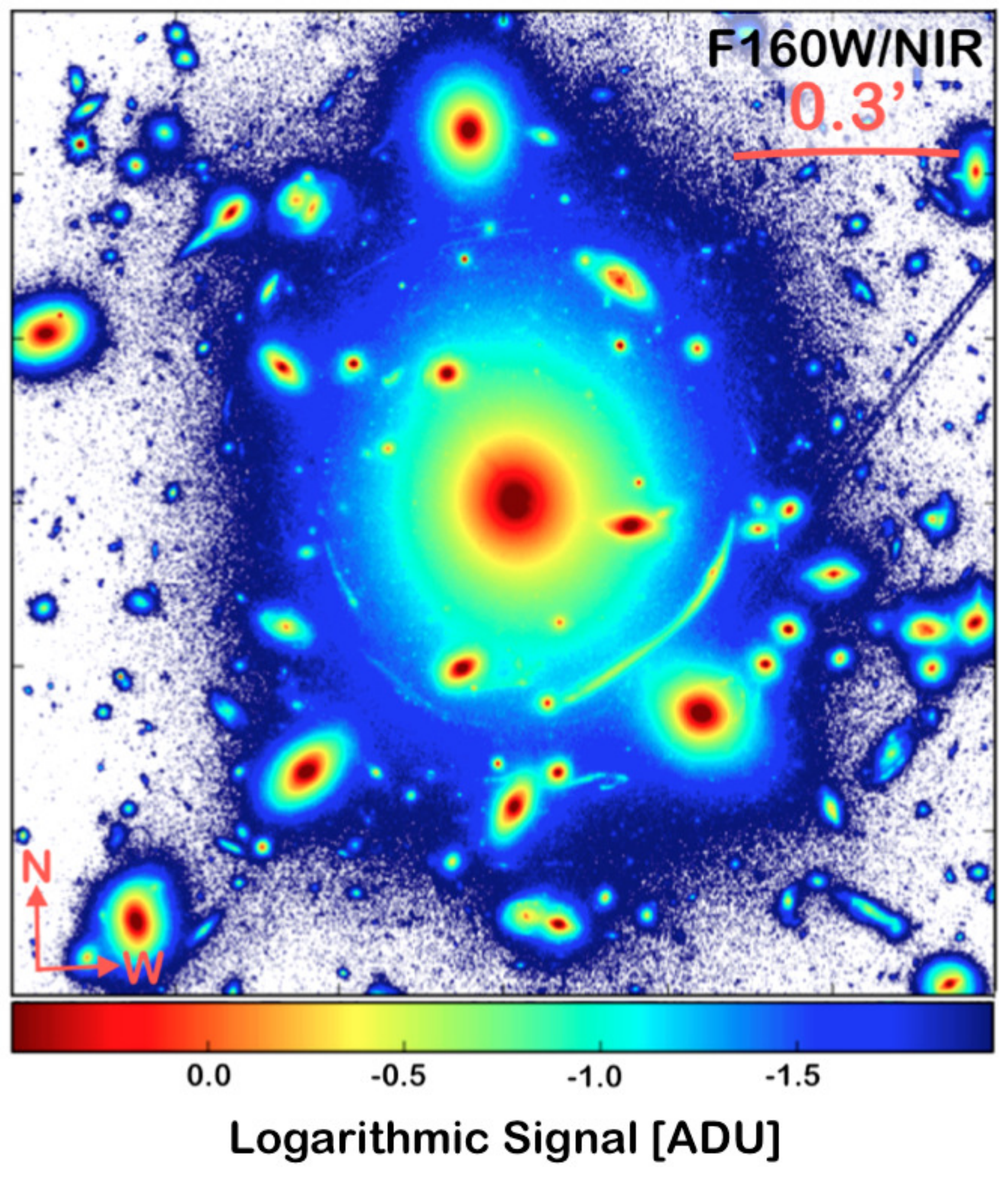}
\caption{Example of the typical ICL+Background signal contaminating the colors of the galaxies in the F160W/NIR image within the cluster Abell-383 (z$_{s}$=0.187).}
\label{ICLexample}
\end{center}
\end{figure}

As discussed through this work, standard aperture photometry on massive cluster fields does not provide as accurate photometric redshifts as expected from field galaxy simulations, where the only source of uncertainty is assumed to be the photometric noise from images. Unlike field samples where galaxies are mostly isolated (apart from pairs, merging systems or projected neighbors) over an almost flat background, galaxies in the cores of massive cluster fields are immersed in a fluctuating background signal mainly dominated by the brightness of the Brightest Cluster Galaxies (BCGs) and the Intra-Cluster-Light (ICL). In deep images of very massive galaxy clusters, as the ones acquired for CLASH, this additional signal (BCG+ICL, hereafter BCL) becomes noticeable and if not properly removed from the images (or included in the simulations as an additional uncertainty), it disrupts the real color of galaxies deteriorating the expected performance of photometric redshifts. As the BCL emission is (generally) dominating and (typically) inhomogeneous, the background estimation on images becomes a non-trivial task. This BCL light varies spatially across the image showing both small- and large-scale structure. This fact complicates its modeling and subtraction since very smooth-maps may not account for the signal between close galaxies and highly resolved-maps may over-subtract light from the brightest galaxies. An example of the BCL signal for the galaxy cluster Abell-383 ($z_{s}=0.187$) is shown in Figure \ref{ICLexample}. 

\vspace{0.2cm}

From a practical point of view when performing aperture photometry, once the sources are detected and their corresponding apertures defined, all remaining pixels on an image are automatically assumed to make part of the background. For the case of the CLASH observations, the BCL signal may well be spread over the entire HST/WFC3 Field-of-View (FoV$\sim$1') contaminating a significant fraction of the pixels used to define the sky-level in images. The intensity of this BCL emission increases with wavelength. Therefore, depending on the properties of a specific cluster and the particular passbands it has observed, photometric colors of galaxies embedded in these halos may turn automatically biased. Whereas for the HST/ACS filters in the CLASH images this effect is moderate (but not negligible), in the NIR it becomes especially significant causing an asymmetric noise distribution (with a long tail) toward positive values. This asymmetric excessed-signal, which cannot be explained by any instrumental background, corresponds to the BCL. If not removed from images, it must be considered as an additional source of noise (uncertainty) when estimating expected fluxes from galaxies.  

\vspace{0.2cm}

Based on all the aforementioned facts, the final observed magnitude ($m_{i}$) of a given galaxy in the i$^{th}$ passband on a particular CLASH field, needs to be described as:

\begin{equation}
m_{i}=m^{o}_{i}+\delta m_{i}^{RMS}+\delta m_{i}^{BCL}
\label{mageq}
\end{equation}

where $m^{o}_{i}$ represents the real flux of the galaxy, $\delta m_{i}^{RMS}$ the additional instrumental noise (depending on the total exposure time and filter response) and $\delta m_{i}^{BCL}$ to the (additional) signal from the BCL (depending on each cluster and passband). Not including this additional source of uncertainty when predicting photo-z performance may lead to a severe overestimation of the real photo-z depth of any survey. In this work we suggest an approach to minimize the impact of the BCL signal on our images improving the overall photometry of cluster galaxies.  

\vspace{0.2cm}
 
\begin{figure}
\begin{center}
\includegraphics[width=8.25cm]{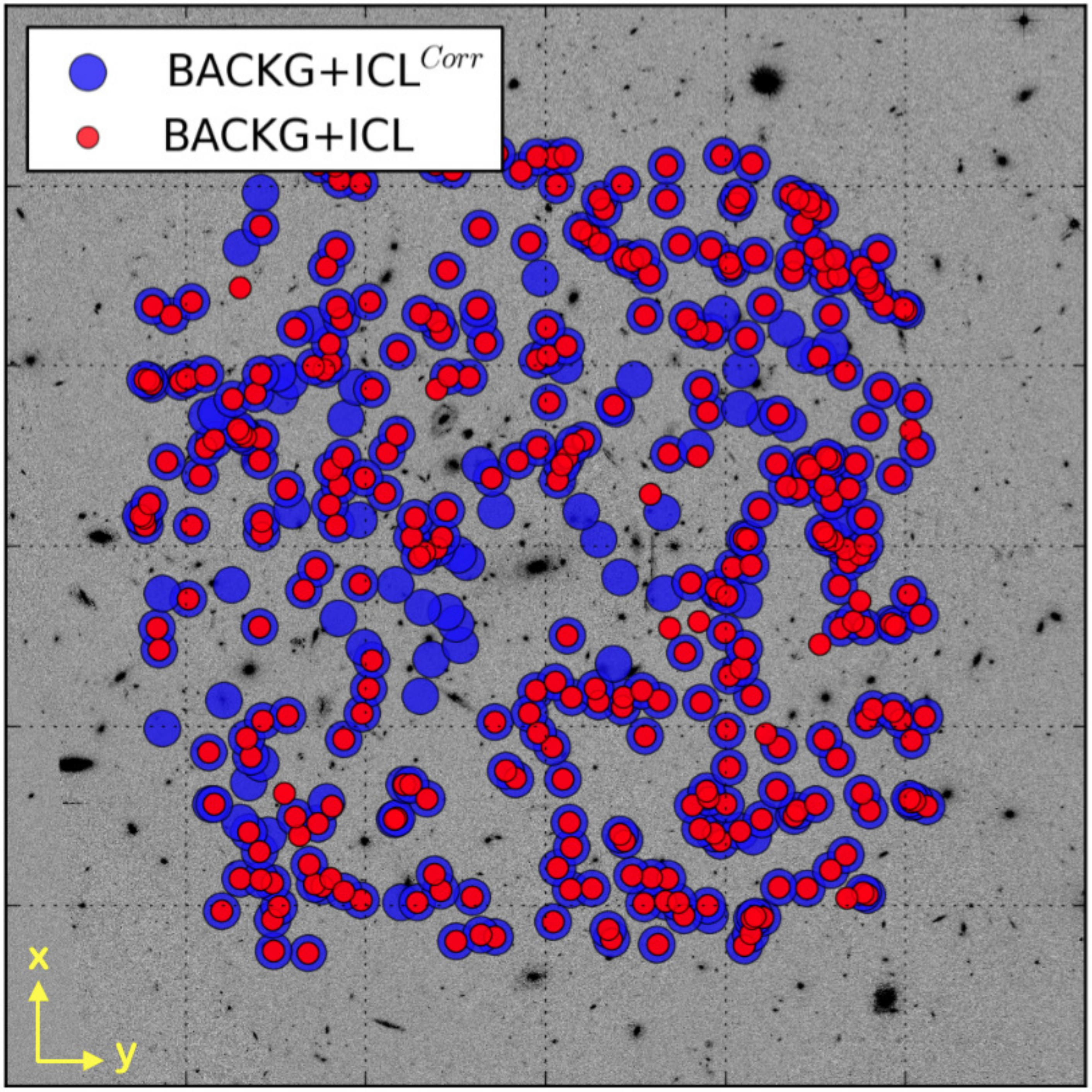}
\caption{The figure illustrates how the detectability of faint sources on dense environments can be compromised due to the ICL+BCG signal. Faint galaxies from the UDF were injected inside our CLASH clusters. The fraction of detected galaxies was  compared before (red dots) and after (blue dots) modeling + subtracting this BCL light. The cleaning processing served as much to increase the detectability of sources as to improve the measured colors of galaxies.}
\label{BCLdetections}
\end{center}
\end{figure}

Another intervening problem when performing photometry on dense environments concerns the detection of faint sources. As already seen in Figure \ref{ICLexample}, innermost regions of massive clusters are strongly dominated by the BCL emission. In certain cases, this signal may be so intense that small and faint galaxies can be completely undetected by \texttt{SExtractor} \citep{1996A&AS..117..393B}. This effect is illustrated in Figure \ref{BCLdetections} where a sample of galaxies from the Ultra Deep Field (UDF; \citealt{2006AJ....132.1729B}) are injected in a CLASH image and the fraction of extracted galaxies by \texttt{SExtractor} is compared when the BCL signal is removed or not. As expected, when the BCL signal is not subtracted (red circles) from images the fraction of retrieved galaxies is significantly smaller than the case when this BCL signal is model and removed (blue circles). Since a large fraction of faint (magnified) galaxies are present in these clusters, this artificial selection effect needs to be seriously considered and fixed. Otherwise, estimations such as luminosity functions for high-z galaxies may be biased as noted in analyses of Frontier Fields clusters (e.g., \citealt{2015ApJ...808..104O}; \citealt{2016arXiv160406799L}; \citealt{2016arXiv161000283B}).

\vspace{0.2cm}

There exists another problem when performing aperture-matched photometry over large wavelength ranges. For the particular case of cluster galaxies, the enormous color indexes between the bluest (UV) and the reddest (NIR) filters makes standard aperture definition very inefficient. The definition of an aperture for a galaxy in a cluster based on a deep NIR band leads to an artificial deterioration of the signal-to-noise in the bluest bands; an effect that directly impacts the overall photometric quality of the survey and eventually underperforms any photo-z estimations.  

\vspace{0.2cm}

Finally, as thoroughly discussed in Sections \ref{photoerr1} \& \ref{photoerr2}, there is another (usually unnoticed) effect impacting the quality of photo-z estimates in dense environments. An inaccurate description of the photometric uncertainties in images, for both bright and faint detections, may cause a significant bias in the redshift distribution of faint background galaxies ($n(z)$) and an artificial deterioration of the photo-z precision for high signal-to-noise galaxies.

\vspace{0.2cm}

Given the complexity of deriving accurate photometry on massive galaxy clusters, the goal of this paper is to propose a new approach to improve the photometry of galaxies in dense environments. This paper is organized as follows: in Section $\ref{observations}$ we describe the CLASH dataset utilized in this work. The pipeline adopted here to derive accurate photometry in clusters is presented through Section $\ref{MBP}$. This includes the definition of new photometric apertures to enhance the signal-to-noise of galaxies in the bluest filters, an efficient subtraction of the ICL from images, a PSF-homogenization of images based on empirical PSF-models and a discussion about the importance of deriving accurate photometric uncertainties when computing photometric upper-limits. Afterwards, the code utilized for photometric redshift estimations is presented in Section $\ref{BPZ}$, including a short discussion about 1. the necessity of including an extra template to fully cover the color-space of galaxies in clusters, 2. the observed zero-point corrections required to match data and models, 3. the spectroscopic redshift sample used to characterize the photo-z estimations, 4. the final performance obtained for galaxies in clusters and 5. the impact of inaccurate photometric uncertainties when computing photo-z estimations for high signal-to-noise galaxies. Finally, Section $\ref{catalogues}$ is devoted to the description of the photometric redshift catalogue and Section $\ref{summary}$ to the discussion of the final results and conclusions.

\vspace{0.2cm}

Unless specified otherwise, all magnitudes here are presented in the AB system. We have adopted the cosmological model provided by the \cite{2014A&A...571A..16P} with parameters $H_{0}$ = 70 km$s^{-1}$ Mpc$^{-1}$ and ($\Omega_{M}$, $\Omega_{\Lambda}$, $\Omega_{K}$) = (0.315, 0.673, 0.00).

\section{Observations}
\label{observations} 

\begin{figure}
\begin{center}
\includegraphics[width=9.cm]{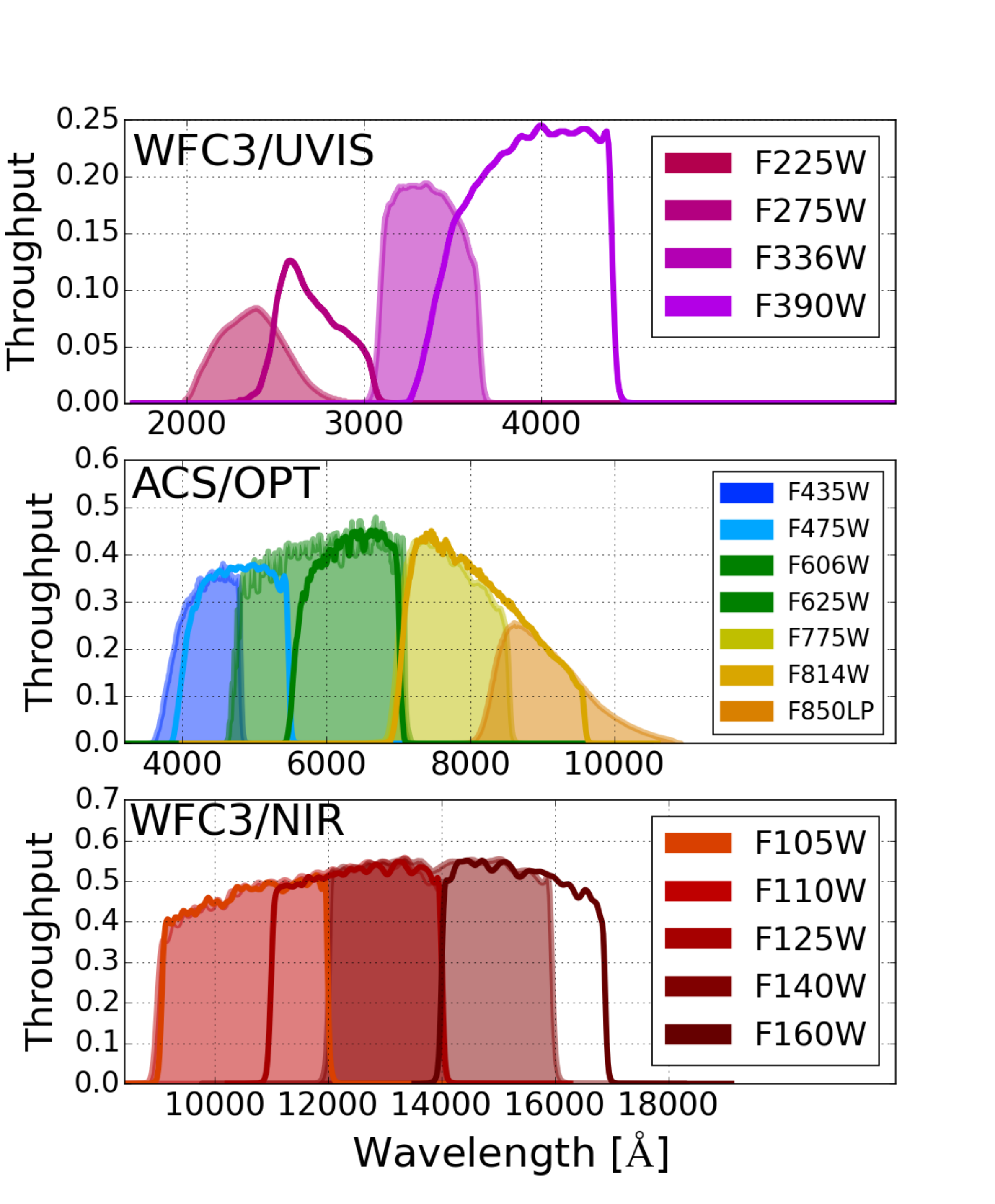}
\caption{The CLASH photometric filter system. It covers the whole UV+Optical+NIR wavelength range combining 16 broad-bands (BB) from 2 different cameras: 4 BB from WFC3/UVIS (top), 7 BB from ACS/WFC (middle) and 5 BB from WFC3/NIR (bottom).}
\label{clashfiltersys}
\end{center}
\end{figure}

CLASH is a Multi-Cycle Treasury programe awarded with 524 $HST$ orbits to image the cores of 25 massive galaxy clusters at intermediate redshifts (see table $\ref{CLASHGCS}$). The observations made use of both the Wide Field Camera 3 (WFC3; \citealt{2008SPIE.7010E..1EK}) and the Advance Camera for Surveys (ACS; \citealt{2003SPIE.4854...81F}) on-board HST, as illustrated in Figure $\ref{clashfiltersys}$. An optimized photometric filter system was selected for the estimation of photometric redshifts, composed by 16 overlapping broad-bands, spanning a wavelength range from the near-ultraviolet (2000$\AA$) to near-infrared (17000$\AA$): 4 filters from the WFC3/UVIS, 5 from WFC3/IR camera and 7 from ACS/WFC. With an averaged exposure time of $\sim$2500 sec (1-2 orbits) per image (or 20 orbits per cluster if all filters are included), the observations reach a typical photometric depth of F814W=28.0 or F160W=26.5 (S/N$>$3). 

\vspace{0.2cm}

Image reduction, alignment and co-adding was done using the \texttt{MosaicDrizzle} pipeline (\citealt{2003hstc.conf..337K}; \citealt{2011ApJS..197...36K}), where a final scale of 0.065 "/pixel was chosen for all the fields. Weight- \& RMS-maps were also computed and utilized during the photometric extraction of sources and the local estimation of noise in images. The reduced images and weight-maps are available at MAST.\footnote{\textcolor{blue}{https://archive.stsci.edu/prepds/clash/}}.

\begin{table}
\caption{The CLASH galaxy cluster sample}
\begin{center}
\label{CLASHGCS}
\begin{tabular}{|l|c|c|c|c|c|c|c|}
\hline
\hline
Cluster  &   $<z_{s}>$ &    R.A.      &      DEC     \\
             &                &  [deg/J2000]  &   [deg/J2000]  \\ 
\hline
\hline
Abell 383	             & 0.187 &  42.0141  &   -03.5293 \\
Abell 209	              & 0.209 & 22.9689   &  -13.6112  \\
Abell 1423              & 0.213 & 179.3223 & 33.6109  \\
Abell 2261              & 0.224 & 260.6134   &  32.1324  \\
RXJ2129+0005      & 0.234 & 322.4165 & 00.0892  \\
Abell 611	               & 0.288 & 120.2367 & 36.0566 \\
MS2137-2353         & 0.310 & 325.0631	&  -23.6611 \\
RXJ1532.8+3021    & 0.345 & 233.2241 & 30.3498 \\
RXJ2248-4431        & 0.348 & 342.1832	 & -44.5309 \\
MACSJ1931-26       & 0.352 & 292.9561 & -26.5758 \\
MACSJ1115+0129  & 0.352 & 168.9663 & 01.4986 \\
MACSJ1720+3536 & 0.387 & 260.0698	& 35.6073 \\
MACSJ0429-02      & 0.399 & 67.4000 & -02.8852 \\
MACSJ0416           & 0.397 & 64.0356 & -24.0733   \\
MACSJ1206-08     & 0.439  & 181.5506 & -08.8009 \\
MACSJ0329-02     & 0.450 & 52.4232 &  -02.1962 \\
RXJ1347-1145       & 0.450 & 206.8776 & -11.7526 \\
MACS1311            & 0.494 & 197.7575 & -03.1777 \\
MACSJ1423.8+2404 & 0.545 & 215.9490 & 24.0793 \\
MACSJ1149          & 0.544 & 177.3980 & 22.3980 \\
MACSJ0717         & 0.548 & 109.3880 & 37.7493   \\
MACSJ2129+0005  & 0.570 & 322.3600 & -07.6923 \\
MACSJ0647         & 0.584 & 101.9620 & 70.2481 \\
MACSJ0744+39   & 0.686 & 116.2200 & 39.4574 \\
CLJ1226+3332    & 0.890 & 186.7427 & 33.5468 \\
\hline
\hline
\end{tabular}
\end{center}
\end{table}  

\section{Multiband Photometry.}
\label{MBP}

This section is devoted to the explanation of how a multi-band aperture-matched PSF-homogenized photometry has been performed on all 25 clusters. In particular, in Section $\ref{RestrictedApertures}$ we discuss the convenience of adopting a different set of apertures with respect to the ones typically utilized for photo-z estimations, to improve the photometry of cluster galaxies by enhancing the signal-to-noise at the shortest wavelengths. In Section $\ref{PSFhomo}$ the adopted approach to generate PSF-homogenized images across filters is introduced. Section $\ref{iclsubt}$ describes the methodology applied to remove the ICL from our images improving the overall photometric quality. Finally, a precise recalibration of the photometric uncertainties along with the estimation of accurate photometric upper-limits for our photo-z estimations is discussed through Sections $\ref{photoerr1}$ \& $\ref{upperlimits}$.  

\subsection{Aperture photometry on clusters.}    
\label{RestrictedApertures}
\begin{figure}
\begin{center}
\includegraphics[width=8.5cm]{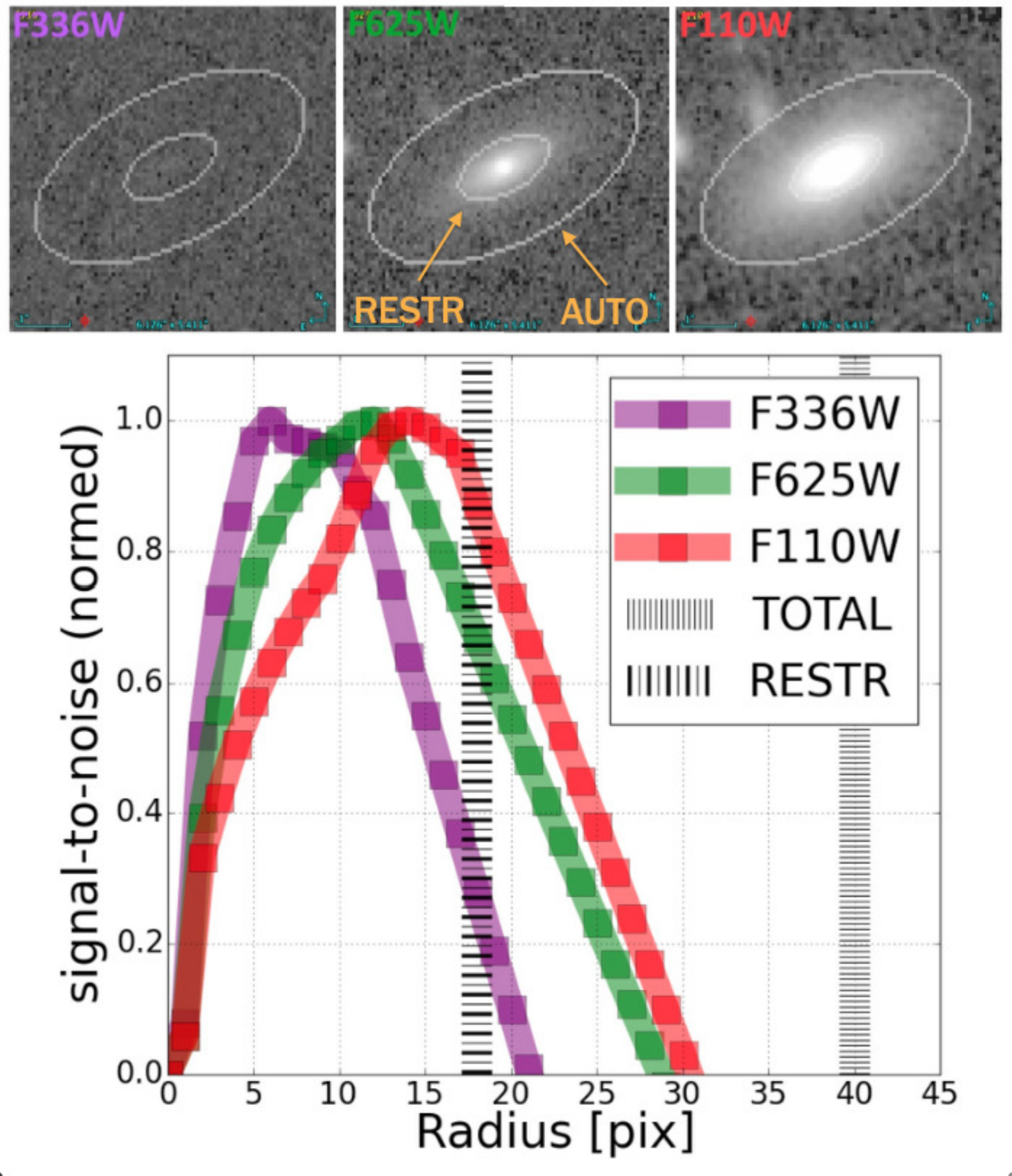}
\caption[Total Restricted Apertures]{Photometric Apertures (I): Upper panel shows an example of an inefficient photometric aperture for an early-type galaxy. Apertures defined based on deep NIR images are systematically much larger than the galaxies in the shortest wavelengths. This effect artificially reduces the signal-to-noise in those bands what has a direct impact on the photo-z estimations. The effect is illustrated in the lower panel where the signal-to-noise as a function of the aperture radius is plot for an early-type galaxy in three different bands (UV/purple, Optical/green and NIR/red). Whereas standard \texttt{SExtractor\_AUTO} (total) apertures include all light from galaxies, they provide a reduced signal-to-noise in the shortest wavelengths. The effect can be mitigated adopting total \texttt{restricted} apertures which provide more accurate colors for photo-z estimations.}
\label{apers1}
\end{center}
\end{figure}

Standard aperture-matched photometry on massive galaxy clusters (mainly dominated by early-types) does not provide as accurate photo-z estimations as for field galaxies. When the Spectral Energy Distribution (SED) of an early-type galaxy is simultaneously observed from the UV to the NIR, it shows large color indexes ($\Delta m>$5 magnitudes). In the case of the CLASH observations, cluster galaxies practically vanish in the shortest wavelengths (UV) while still preserving a high signal-to-noise in the reddest (NIR) filters. When a photometric aperture is defined for these galaxies according to a deep NIR image and the aperture is transported to images with shorter wavelengths (shorter than the $\lambda<4000\AA$-break at rest-frame), the so-defined apertures are systematically much larger than the entire galaxies. These ill-defined apertures artificially reduce the signal-to-noise of galaxies in the bluest filters (especially in the UV), making the photometry enormously noisy and uncertain. This issue is illustrated in the upper panel of Figure $\ref{apers1}$, where a \texttt{SExtractor}\_AUTO aperture is overlaid on top of the galaxy in three different filters (F336W/UV, F625W/OPT $\&$ F110W/NIR). It is worth emphasizing that this effect has nothing to do with differences in the PSF among images. It is a rather specific problem of early-type galaxies observed through a large wavelength range (as the one adopted in CLASH). The fact that early-type galaxies usually represent a subdominant population in deep field galaxy surveys (mainly dominated by late-type galaxies with moderate color indexes), explains why standard aperture photometry generally yields more accurate photo-z estimations than those obtained specifically in massive cluster fields.

\vspace{0.2cm}
   
In order to circumvent this situation and be able to derive an enhanced photometry for cluster galaxies in the bluest filters, we adopted a new set of photometric apertures. On the one hand, we define total \texttt{restricted} apertures by forcing \texttt{SExtractor} to define total (\texttt{AUTO}) magnitudes with the smallest radius possible; i.e., setting the \texttt{SExtractor PHOT\_AUTOPARAMS}\footnote{The \texttt{PHOT\_AUTOPARAMS} serves to regulate the definition of an elliptical aperture around every detection.} parameter to a value of \texttt{1.0,1.0}. These \texttt{restricted} apertures have the advantage of integrating most of the light from the galaxies while keeping a higher signal-to-noise than the standard \texttt{SExtractor\_AUTO} magnitudes. This effect is illustrated in the lower panel of Figure $\ref{apers1}$, where the S/N of a galaxy within two photometric apertures (\texttt{AUTO} \& \texttt{restricted}) is compared in three different filters. We preferred to adopt these \texttt{restricted} apertures rather than the standard \texttt{SExtractor\_ISO} magnitudes since the latter are more sensitive to small variations in the PSF across images (see discussion in Section \ref{PSFhomo}). 

\vspace{0.2cm}

\begin{figure}
\begin{center}
\includegraphics[width=8.cm]{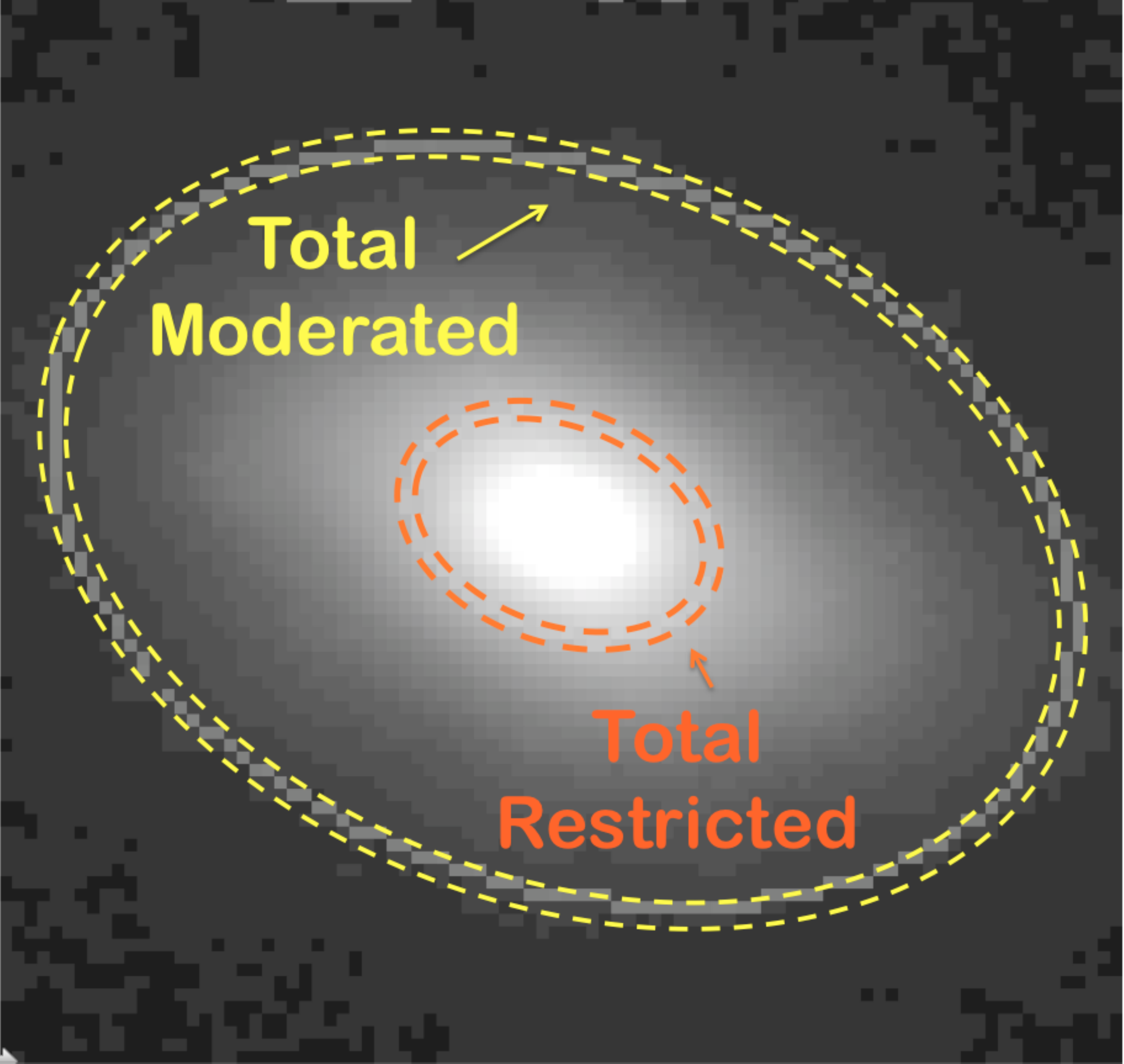}
\caption[Total Moderated Apertures]{Photometric Apertures (II): in order to be able to derive unbiased physical properties from the galaxies (such as stellar masses, ages or metallicities), we also defined total \texttt{moderate} apertures to integrate (almost) all light from the galaxies while reducing potential contamination from neighbors. The figure compares the double-set of photometric apertures adopted in this work.}
\label{apers2}
\end{center}
\end{figure}

On the other hand, in order to integrate all light from galaxies and be able to derive unbiased physical properties (such as stellar masses, ages or metallicities), we decided to include a secondary set of apertures. In this case, we defined total \texttt{moderate} apertures by forcing \texttt{SExtractor} to define total (\texttt{PETRO}) magnitudes with apertures not larger than the distance at which the signal-to-noise of galaxies drops to zero in the detection images; i.e., setting the \texttt{SExtractor PHOT\_AUTOPARAMS} parameter to a value of \texttt{2.0,1.0}. These apertures, similar to the standard \texttt{SExtractor\_AUTO} magnitudes but slightly smaller, served to integrate (almost) all light from the galaxies while reducing potential contaminations from neighboring galaxies. Figure $\ref{apers2}$ shows an example of both sets of apertures (i.e., \texttt{restricted} \& \texttt{moderate}) adopted in this work.

\vspace{0.2cm}
 
As explained in Section $\ref{catalogues}$, the final photometric catalogue includes both types of photometries derived on the 16 bands. In this work, total \texttt{restricted} magnitudes were used for photo-z estimations (Section $\ref{BPZ}$) whereas total \texttt{moderate} magnitudes will be applied in a separate paper (Molino et al., in prep.) to derive physical properties of cluster galaxies (such as ages, metallicities, extinctions or stellar masses). 

\subsection{PSF Homogenization}
\label{PSFhomo}
In order to deal with the differences in the PSF among filters and derive an accurate aperture-matched photometry, we decided to homogenize the whole set of images to a common PSF value. For the case of CLASH, we chose to bring it to the broader PSF condition given by the WFC3/IR camera. To do so, we relied on the IRAF \texttt{psfmatch} routine (\citealt{1995ASPC...77..297P}) to compute the convolution between different images. Basically, this routine computes an empirical kernel between two PSF-models (one model from the original image to be degraded and one model for the final PSF condition to be reached) previous to the convolution process. 

\vspace{0.2cm} 

\begin{figure}
\begin{center}
\includegraphics[width=7.5cm]{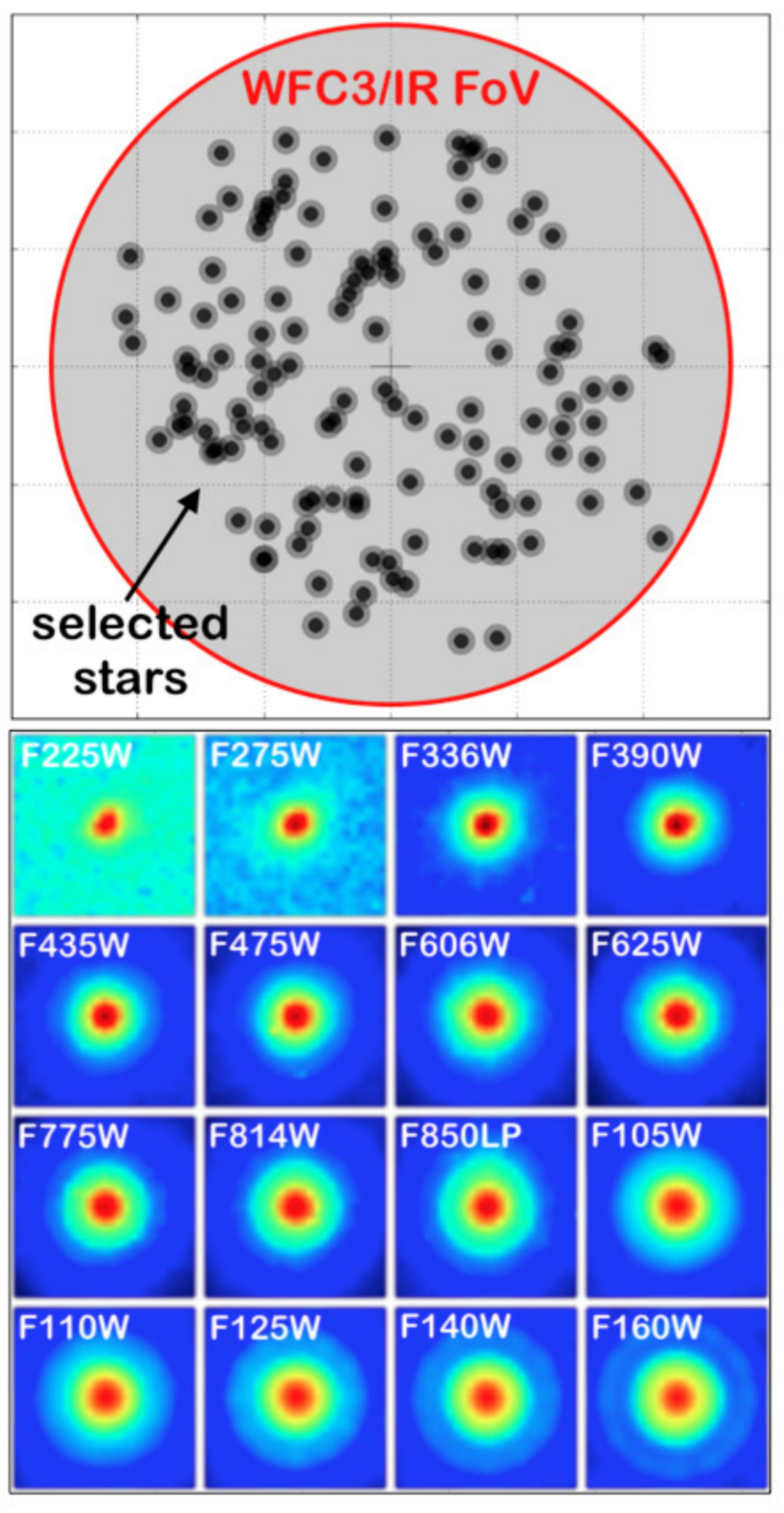} 
\caption{To perform accurate aperture-matched photometry across bands, we decided to bring all images to a common PSF condition using the \texttt{psfmatch} routine from IRAF. To do so, we selected a total of 131 stars within the WFC3/IR FoV from different clusters (top) to derived empirical PSF-models for every filter (bottom).}
\label{PSFmodels}
\end{center}
\end{figure}

In order to execute the \texttt{psfmatch} routine, it was necessary to generate PSF-models for every individual image\footnote{These PSF-models are available at the following website: \textcolor{blue}{https://archive.stsci.edu/prepds/clash/}}. To achieve this goal, we carefully scanned the 25 clusters seeking for high signal-to-noise stars. Each star was then double-checked to assure that it was neither photometrically saturated nor too close to another bright neighbor within a 25-pixel square box around the star (i.e., within the PSF-model grid-size). Finally, the remaining sample of 131 stars was combined and normalized in every band to produce the PSF-models shown in the lower panel of Figure \ref{PSFmodels}. A certain spatial and temporal PSF variability is expected for the $HST$ images, however since the IRAF \texttt{psfmatch} routine cannot handle more than a PSF-model per image, we opted to build empirical ``averaged'' models combining stars from different locations (always within the HST WFC3/IR FoV) and from different epochs rather than using position-dependent models (like Tiny-Tim; \citealt{1993ASPC...52..536K}). Although a $\delta PSF<3\%$ scatter was observed based on this compilation of stars, the photometric apertures adopted in this work (Section \ref{RestrictedApertures}) should be less sensitive to these small position-dependent differences. 
 
\subsection{ICL subtraction.}
\label{iclsubt}

Galaxies within the cores of massive cluster fields are immersed in a fluctuating background signal mostly dominated by the brightness of the BCG and the ICL. If this light is not properly removed from images, it may seriously disrupt the color of the galaxies deteriorating the performance of photo-z estimations. In order to diminish this nuisance effect, in this work we decided to remove this additional signal from our images previous to the computation of the photometry. To do so, we initially started using \texttt{SExtractor} to derived background-maps directly from our images which would be eventually subtracted from the former\footnote{If requested \texttt{SExtractor} can compute background-maps from images (\texttt{CHECKIMAGE\_TYPE}=\texttt{BACKGROUND}). These maps can be used afterwards to generate background-free images.}. The fact that the BCL light shows both large and small-scale structure made its modeling complicated since very smoothed-maps may not account for the signal between close galaxies and highly resolved-maps may over-subtract light from the brightest galaxies. After testing different configurations for \texttt{SExtractor}, we concluded that in order to effectively remove the BCL signal from the galaxies, we need to use aggressive background configurations (deriving high resolution maps). However this approach turned out not to be ideal since most light from the brightest galaxies was unwillingly removed, biasing the photometry of a large fraction of the galaxies. 

\vspace{0.2cm}  
 
\begin{figure}
\begin{center}
\includegraphics[width=8.5cm]{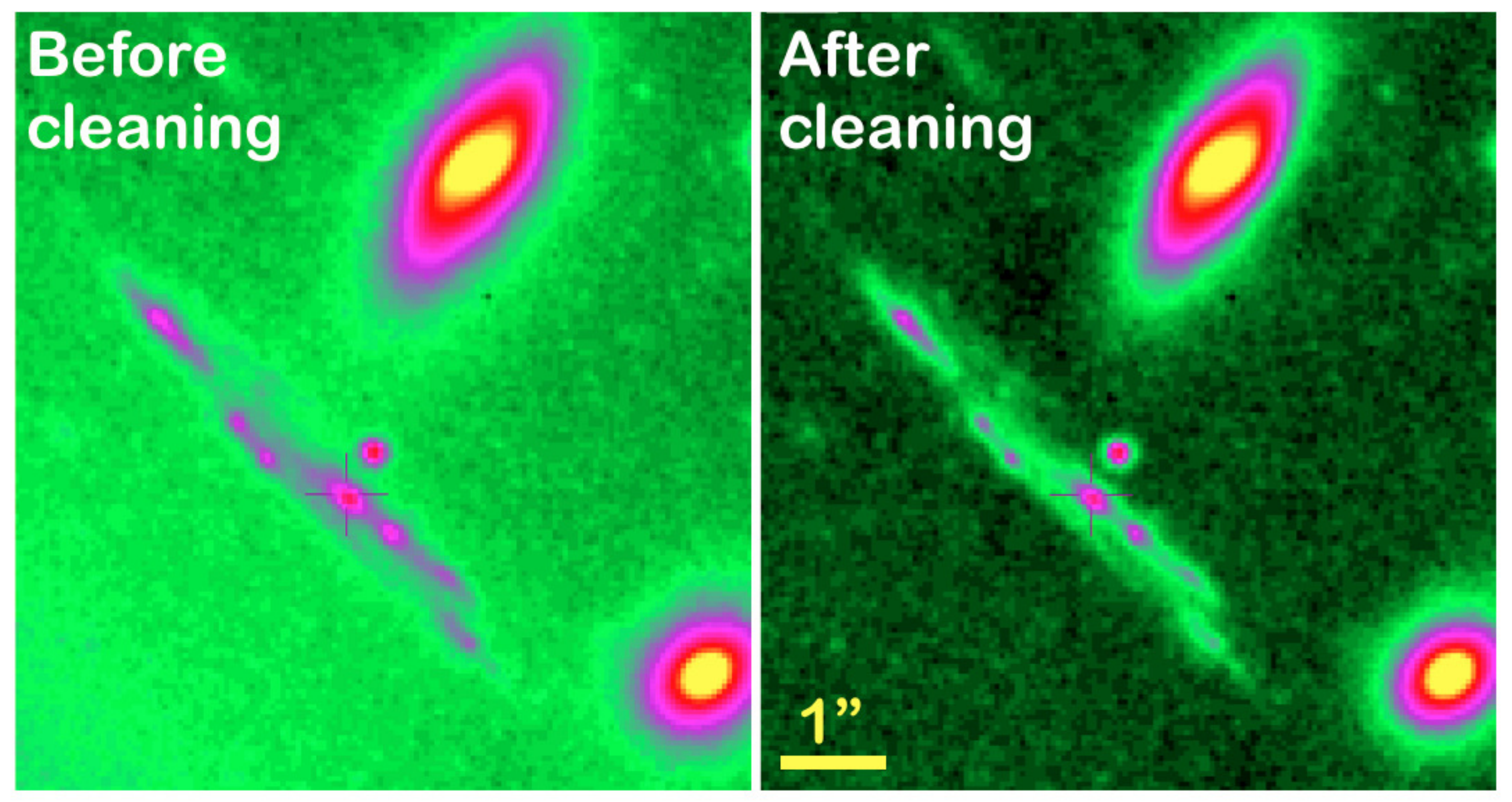} 
\caption{In this work we utilized the \texttt{CHEFs} software to derive multi-wavelength ICL-maps per cluster. These maps were removed from the images before performing the photometry. The figure shows an example of this ICL cleaning for a group of small galaxies in the A209 cluster.}
\label{cleaningICL}
\end{center}
\end{figure}

For this reason, we decided to rely on the \texttt{CHEFs} software (\citealt{2012ApJ...745..150J}; \citealt{2015MNRAS.453.1136J}; \citealt{2016ApJ...820...49J}) instead which is one of the state-of-the-art codes for galaxy modeling. Basically, this software utilizes a library of Chebyshev-Fourier mathematical functions in a non-parametric fashion to efficiently model the light surface distribution of galaxies irrespective of their morphologies (we refer the interested reader to the aforementioned papers for more details about the software). In order to provide the \texttt{CHEFs} with a complete list of sources to model (and remove) in every image, we initially ran \texttt{SExtractor} on the deep NIR detection images using an aggressive background configuration to detect as many galaxies as possible. Based on the resulting source catalogue, afterwards we ran the \texttt{CHEFs} on every science image to model and subtract every detection. This process served to compute ICL-maps per image and cluster\footnote{These multi-wavelength ICL-maps will be presented in a separate paper}. Finally, these ICL-maps were subtracted from the original images deriving ``background-free" images to be used for the final photometry (Section \ref{RestrictedApertures}). Figure \ref{cleaningICL} shows an example of this ICL cleaning for a group of small galaxies in the A209 cluster. 

\subsection{Photometric Uncertainties (I)}
\label{photoerr1}

It is known that \texttt{SExtractor} systematically underestimates the photometric uncertainties of sources due to the fact that science images (usually) undergo several processing steps (dithering, degradation, stacking, registration, etc), which introduces correlations between neighboring pixels. This correlation makes the background noise different from a Poissonian distribution and so the \texttt{SExtractor} uncertainties no longer accurate (see \citealt{2014MNRAS.441.2891M} for an extended discussion). Besides the estimation of the instrumental noise in images, as stressed in Section \ref{intro}, the original colors of galaxies embedded in massive clusters are expected to be altered due to the presence of the BCL signal. Since this secondary source of ``noise" cannot be accounted by \texttt{SExtractor}, the combination of both effects may lead to a severe underestimation of the real photometric uncertainties.    

\vspace{0.2cm}

In order to retrieve a robust photometry for CLASH, we decided to explore up to what extend the uncertainties reported by \texttt{SExtractor} were accurate and how much these uncertainties may be reduced after the modeling and removal of the ICL signal from images. To tackle these questions, we followed a similar approach as that presented in Section \ref{intro} injecting galaxies from the $UDF$ in our images and quantifying how much the original magnitudes of these galaxies would change simply because they were now observed through a different background condition. It is worth noting that the so-derived magnitude variations (input minus output) represent a direct and clean quantification of the real photometric uncertainties taken place on our images. 

\vspace{0.2cm}

\begin{figure}
\begin{center}
\includegraphics[width=8.5cm]{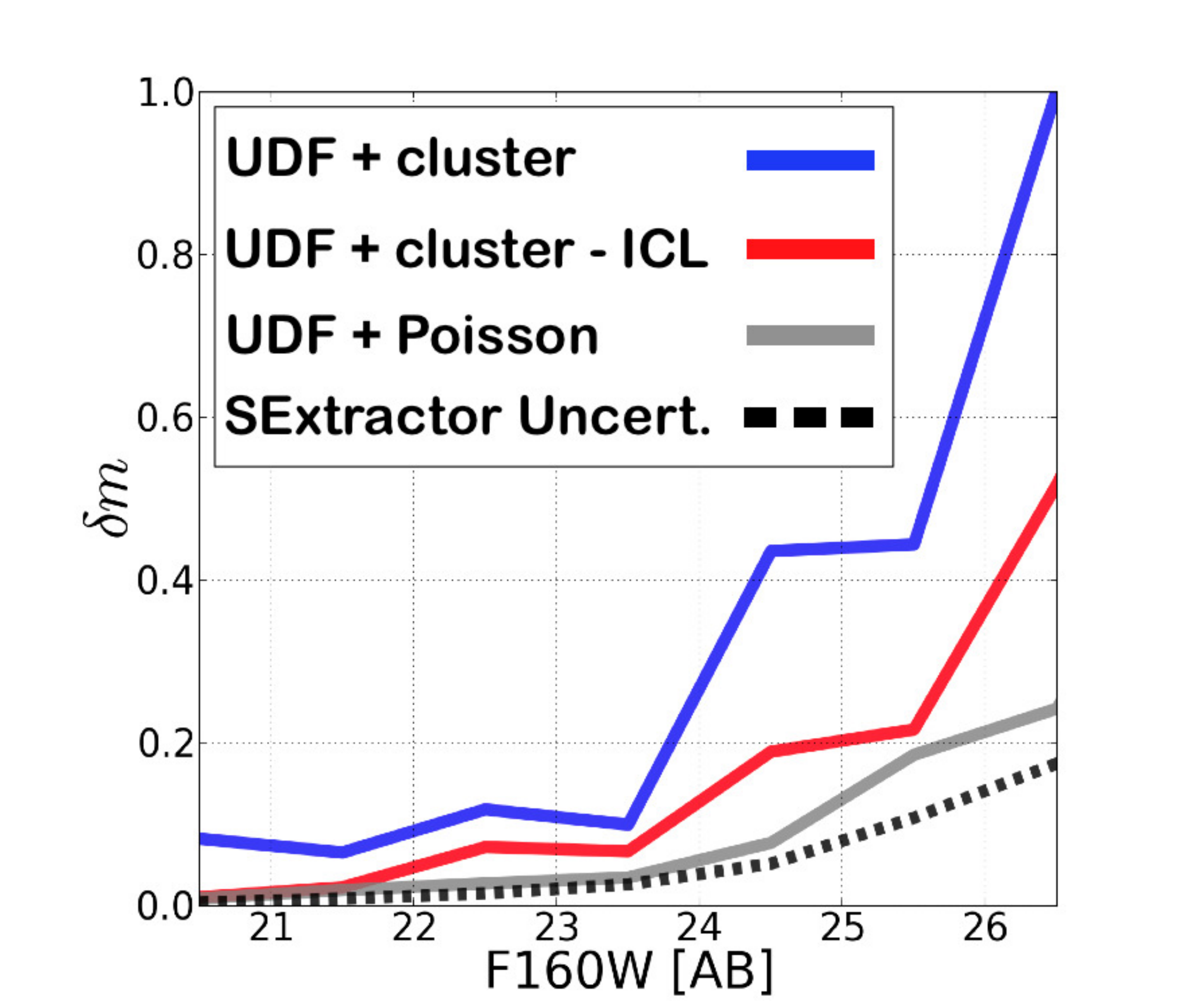} 
\caption{To study the reliability of \texttt{SExtractor} deriving photometric uncertainties, galaxies from the $UDF$ were injected in our images under different background conditions: instrumental background plus BCL (blue), instrumental background minus BCL (red) \& pure Poisson background (grey). The observed variation of their original magnitudes ($\delta m$) as a function of a reference magnitude (F160W) showed that the uncertainties reported by \texttt{SExtractor} were only accurately described for the Poisson case, but they were clearly underestimated for the CLASH images. As expected, the ICL subtraction (red) quantitatively improved the quality of our original photometry (blue).}
\label{backunc}
\end{center}
\end{figure}

Initially, we confirmed that the uncertainties reported by \texttt{SExtractor} were accurate for images with a background noise well-described by a Poisson distribution. To do so, we utilized the \texttt{mknoise} routine from IRAF to background-scaled the $UDF$ image to the level of the CLASH depth by adding Poisson noise; i.e., making the original magnitudes from the $UDF$ galaxies as noisy as the galaxies in the CLASH fields. On this new image, we ran \texttt{SExtractor} in dual-image mode using the original $UDF$ image for detections (to prevent changes in the magnitudes due to differences in the apertures), and compared the variation in the \texttt{SExtractor} total (MAG\_AUTO) magnitudes ($\delta_{m}$) with the reported photometric uncertainties (MAGERR\_AUTO). As seen in Figure \ref{backunc}, a good agreement was found between uncertainties (dashed black line) and empirical magnitude variations (grey solid line). 

\vspace{0.2cm}
 
Later on, we repeated the same exercise but now injecting the $UDF$ galaxies in both the original and the new ``background-free" CLASH images (see Section \ref{iclsubt}). As seen in Figure \ref{backunc}, the photometric bias induced by the clusters (blue line) was always much larger than the uncertainties reported by \texttt{SExtractor}. Although subtracting the ICL signal served to quantitatively mitigate this difference (red line), in both cases the reported uncertainties were severely underestimated. 

\vspace{0.2cm}

In the light of the previous results, we decided to rely on an empirical approach similar to those followed by \citealt{2000AJ....120.2747C};  \citealt{2003AJ....125.1107L}; \citealt{2006ApJS..162....1G}; \citealt{2007AJ....134.1103Q}; \citealt{2014MNRAS.441.2891M} or \citealt{2016arXiv160906919N} to correct the photometric uncertainties. Basically,  the idea behind these methods is the following: every detected object by \texttt{SExtractor} in the detection image is masked out using the \texttt{SExtractor} segmentation-map. Then $\sim$50.000 apertures are thrown over the remaining (blank) area, measuring both the enclosed signal and the RMS inside it. The procedure is repeated for apertures in the 1-250 pixel range, correcting appropriately by the total exposure time of the pixels belonging to them using the weight-maps. As a result we observed two effects. Although the CLASH images were accurately described by a Poisson distribution on small scales, on larger scales the RMS start departing from the \texttt{SExtractor} expectations indicating the presence of correlations among pixels. In order to quantify the importance of the ICL in this analysis, we repeated the same exercise on the original images. As expected, in this case we observed an asymmetric background signal enclosed within the apertures, becoming specially intense for the reddest filters (NIR). This long tail of positive values caused by the BCL, was making the noise distribution in images no longer Gaussian.  

\subsection{Photometric Upper-limits}
\label{upperlimits}

Photometric upper-limits represent the minimum detectable signal from an astronomical image. These magnitudes are important pieces of information when computing photo-z estimations, serving as much to constrain the color-redshift space (i.e., narrowing the posterior redshift probability distribution function) as to reduce the fraction of catastrophic outliers (Section \ref{BPZ}). By construction, an upper-limit depends on the adopted photometric aperture ($A$), on the noise properties of images ($\sigma_{rms}$) and on the significance for the detection to be considered real. These parameters are related as shown in the following equation:

\vspace{0.1cm}
  
\begin{equation}
\label{upperlims1}
\,\,\,\,\,\,mag^{n-\sigma}_{upp} = {-2.5\,\times\,log(n \times \sigma_{rms})\,+zp_{i}}
\end{equation}

\vspace{0.1cm}

where $\sigma_{rms}$ denotes the 1-sigma interval estimated from the noise distribution within a given aperture, $n$ the number of requested sigmas for the limiting magnitudes and $zp_{i}$ the photometric zero-point. 

\vspace{0.2cm}

Taking into account that upper-limits are noise-dependent estimations, we investigated how inaccurate descriptions of photometric uncertainties might affect its definition. Based on equation \ref{upperlims1}, we initially compared the expected differences in magnitude for our upper-limits if they were derived assuming a Poisson-like behavior for the background-noise (as \texttt{SExtractor} does) or using a rather empirical estimation of it (see Section $\ref{photoerr1}$). To compute these quantities, we calculated first the typical sizes of non-detected galaxies in the bluest filters. To do so, we ran \texttt{SExtractor} in dual-image mode on the $F336W$, $F390W$ \& $F435W$ images using the corresponding NIR-detection image for both detections and aperture definitions. We observed that galaxies with magnitudes m=99. (i.e., non detected) had typical sizes of 10-70 pixels, with an average value of $\sim$20 pixels. Converting these sizes into apertures, we found a difference of $\delta$m=0.5 for galaxies as small as 8 pixels or a $\delta$m=1.0 for galaxies as large as 16 pixels. As expected, these differences decreased for smaller apertures since it is precisely on the smallest-scales where the background-noise recovers its Poisson distribution. 

\begin{figure}
\begin{center}
\includegraphics[width=8.cm]{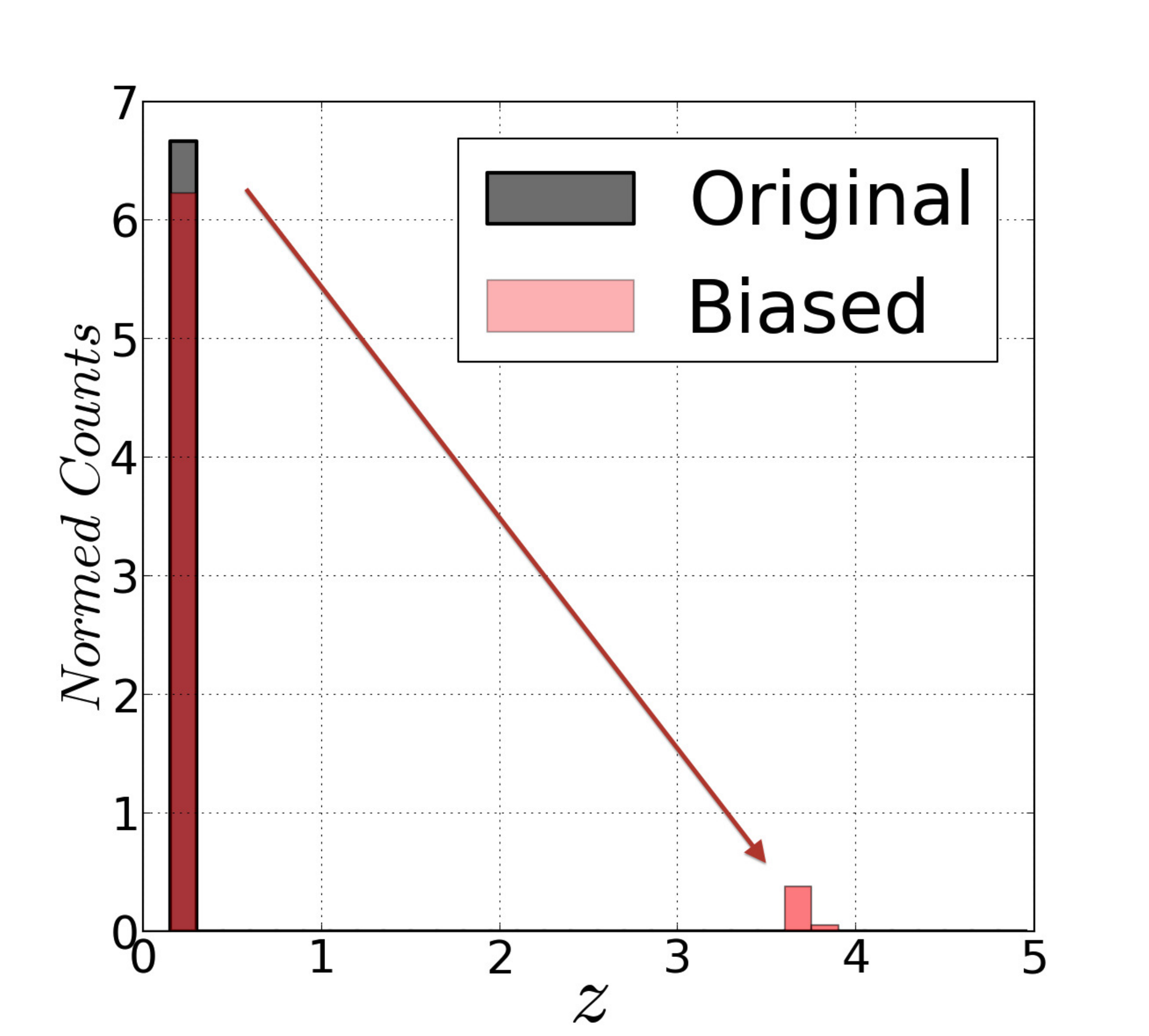} 
\caption{Bias in the $n(z)$ of background galaxies due to inaccurate upper-limits. Mock catalogues for cluster galaxies at different redshifts (vertical grey bar) were generated to explore the impact of faulty upper-limits when computing photo-z estimates. As illustrated in the figure, underestimated photometric errors, leading to overestimated upper-limits in the shortest wavelengths, were forcing \texttt{BPZ} to favor high-z solutions due to a misclassification between the Balmer and the Lyman-break. The bias causes an accumulation of cluster galaxies at z$>$3 (vertical red bar).}
\label{upplimfig1}
\end{center}
\end{figure}

\vspace{0.2cm}

Assuming these numbers to be representative, we designed a set of simulations to understand the impact of these biased upper-limits on our photo-z estimates. To do so, we generated mock catalogues simulating the red sequence of a galaxy cluster at different a redshifts. Each sample, composed by 500 galaxies, was perfectly reproducing as much the colors of the \texttt{BPZ} early-type templates (Section \ref{BPZ}) as the magnitude selection function observed in our images. We perturbed the model magnitudes according to the real noise of images but we set the photometric uncertainties according to  \texttt{SExtractor}. Finally, we derived (biased) upper-limits for non-detected galaxies and re-ran \texttt{BPZ}.

\vspace{0.2cm}

We noticed that these faulty upper-limits were artificially increasing the spectral break of our simulated galaxies, forcing \texttt{BPZ} to favor high-z solutions due to a misclassification between the Balmer (4000$\AA$) and the Lyman ($912\AA$) break. As illustrated in Figure \ref{upplimfig1}, this effect was causing a small fraction of the simulated cluster galaxies (vertical grey bar) to be shifted at z$>$3.0, leading to an artificial excess (small red peak) in the redshift distributions $n(z)$. The fraction of misclassified galaxies varied from 2\% to 4\% for galaxies with a $\delta$m=0.75 magnitudes or $\delta$m=1.0, respectively. It is worth noting that this artificial peak had already been reported in previous catalogues but the source of this signal was unclear. The new definition of upper-limits adopted in this work dramatically mitigates this problems, retrieving more accurate distribution of background galaxies. Based on what has been stated before, we conclude that this effect should not be ignored for massive cluster fields as those of CLASH. Photo-z estimates for non-detected faint galaxies due to either the ICL or the proximity to a bright neighbor, may be severely biased if their magnitudes are replaced by faulty upper-limits. 

\section{Bayesian Photometric Redshifts.}
\label{BPZ}
\begin{figure}
\begin{center}
\includegraphics[width=8.5cm]{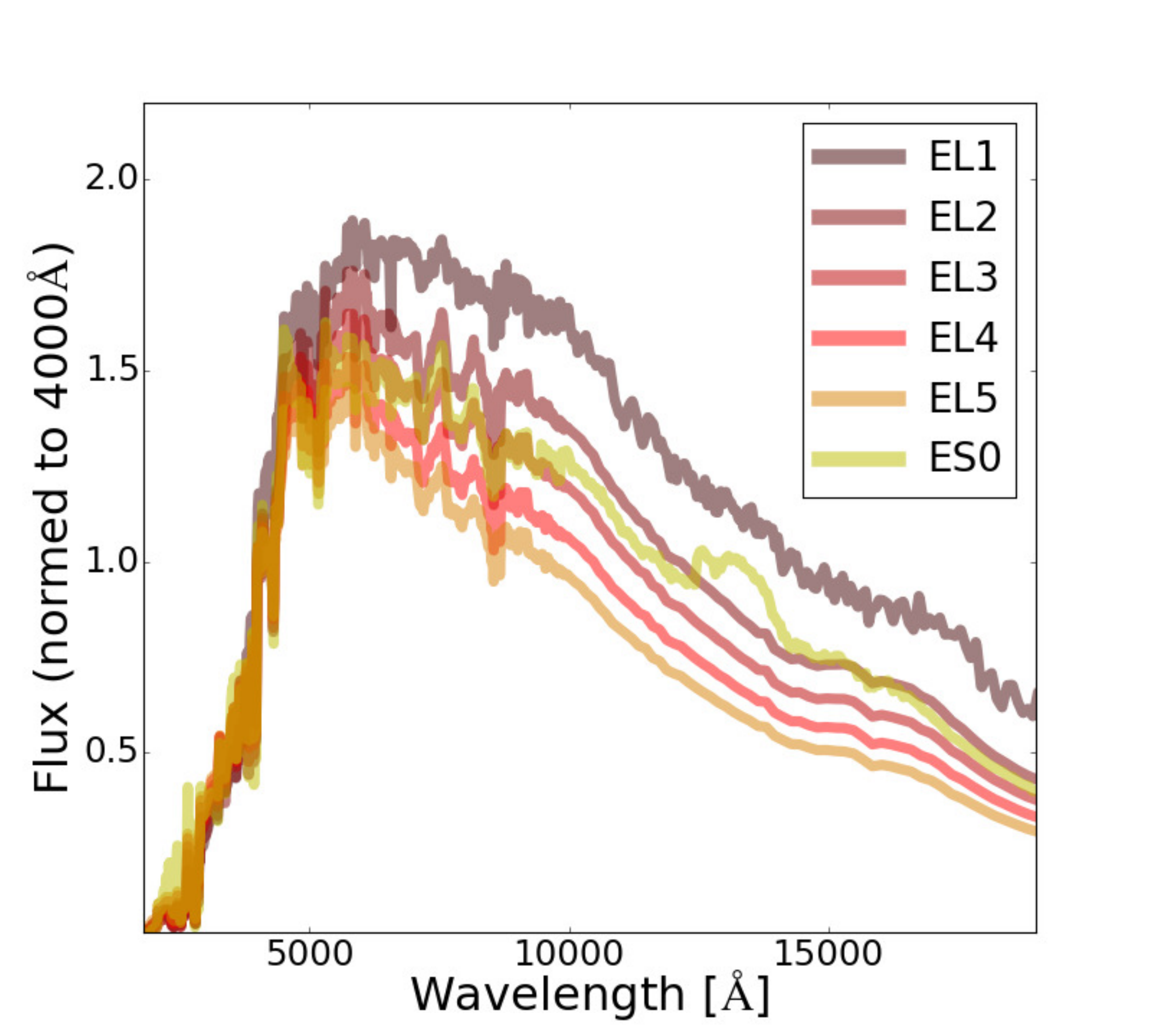} 
\caption[BPZ SEDs]{SED templates of early-type galaxies utilized in this work. In order to fulfill the color-space of galaxies in clusters, it was necessary to incorporate an additional template (EL1) to the original library of \texttt{BPZ2.0} for very red galaxies.}
\label{sedtemp}
\end{center}
\end{figure}

We calculate photometric redshifts (photo-z) using an updated version of the Bayesian Photometric Redshifts (\texttt{BPZ}) code (\citealt{2000ApJ...536..571B}; \citealt{2006AJ....132..926C}), which includes several changes with respect to its original version (see Molino et al. 2014, for more details). A new library composed of six SED templates originally drawn from Projet d'\'Etude des GAlaxies par Synth\`ese \'Evolutive (PEGASE: \citealt{1997A&A...326..950F}) but then re-calibrated using FIREWORKS photometry and spectroscopic redshifts (\citealt{2008ApJ...682..985W}) to optimize its performance. In addition to these basic six templates, four GRAphite and SILicate ($GRASIL$; \citealt{2005AIPC..761..187P}) and one starburst template have been added. As explained in section $\ref{newtemplate}$, an additional early-type template (EL1 in Figure $\ref{sedtemp}$) was required to fulfill the color-space of the reddest cluster galaxies. Therefore, the library used in this work includes six templates for elliptical galaxies, two for spiral galaxies and four for starburst galaxies along with emission lines and dust extinction. The opacity of the intergalactic medium was applied as described in \cite{1995ApJ...441...18M}. 

\vspace{0.2cm}

The \texttt{BPZ2.0} also includes a new empirically derived prior by the redshift distributions measured in the GOODS-MUSIC (\citealt{2009A&A...504..751S}), COSMOS (\citealt{2007ApJS..172...38S}) and UDF (\citealt{2006AJ....132..926C}) catalogs. This prior has proved to provide excellent results in deep field surveys (\citealt{2014MNRAS.441.2891M}). However, since the cluster galaxies considered in this work are (typically) clustered at certain redshift ranges with peculiar magnitude distributions, we preferred to apply a ``flat" (rather than the standard luminosity-based) prior on both galaxy type and redshift. A new empirical prior optimized for cluster galaxies will be presented in a separate paper. Likewise, the \texttt{BPZ2.0} provides either an estimation of the galaxy stellar mass (obtained by applying the color-M/L ratio relationship established by \citealt{2011MNRAS.418.1587T}) and an estimation of the absolute magnitudes  according to the most likely redshift and spectral-type per each galaxy.

\vspace{0.2cm}

We evaluate the performance of our photo-z estimates using the normalized median absolute deviation (NMAD) since it manages to get a stable estimate of the spread of the core of photo-z distribution without being affected by catastrophic errors making the photo-z error distribution to depart from a pure Gaussian distribution. Along with the scatter from the error distribution, it is also important to control any systematic bias $\mu$ in the redshift distribution and to quantify the fraction of (potential) catastrophic errors. The NMAD is defined as:
  
\begin{equation}
\sigma_{NMAD} = 1.48 \times median({\left | \delta z - median(\delta z) \right | \over 1+z_{s}})
\label{nmadeq}
\end{equation}

being $z_{b}$ the photometric redshift, $z_{s}$ the spectroscopic redshift and $\delta z$=($z_{b}-z_{s}$). In this work, we adopt the following definition for catastrophic outliers:

\begin{equation}
\eta = {\left | \delta z \right | \over 1+z_{s}} > 5 \times \sigma_{NMAD} 
\end{equation}

\vspace{0.2cm}

Among others factors, the photo-z precision depends on the number of filters a galaxy is observed through (\citealt{2009ApJ...692L...5B}). In order to define ``homogeneous photo-z samples'', we selected only those galaxies falling within the area covered by the 16 filters; i.e., within the WFC3/IR FoV. Although this criterium may reduce the survey effective area, it guarantees that the galaxies were observed under similar circumstances (in terms of number of orbits and wavelength coverage). Meanwhile, we decided to exclude the UVIS/F225W \& UVIS/F275W filters when running \texttt{BPZ2.0}. Although these filters might have served to break possible redshift degeneracies, its limited depth was actually worsening as much the overall photo-z precision as the fraction of catastrophic outliers.

\subsection{An extension of the $\texttt{BPZ}$ library of templates.}
\label{newtemplate}

\begin{figure}
\begin{center}
\includegraphics[width=8.5cm]{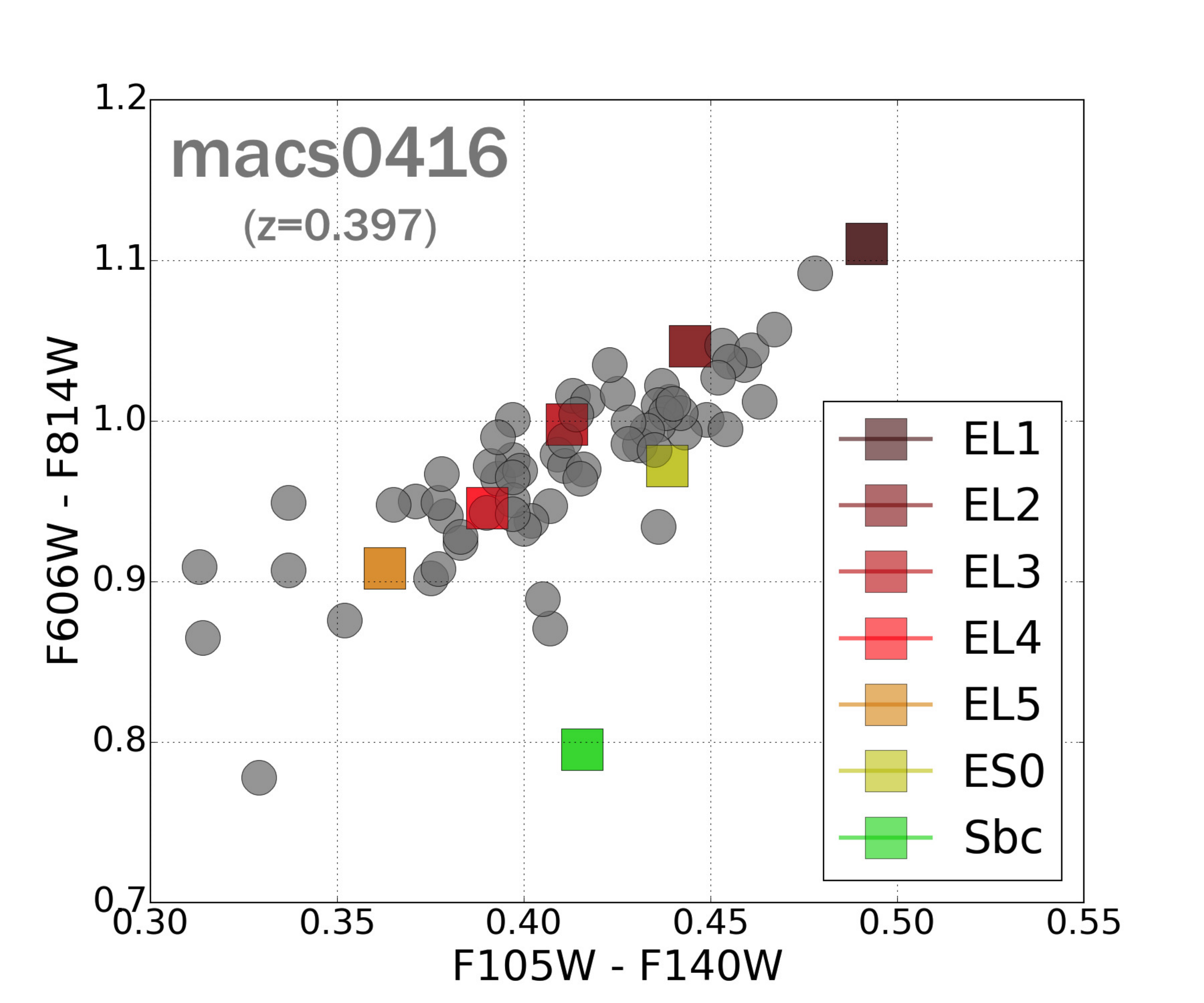}
\caption{Example of the color-color diagram for both the \texttt{BPZ2.0} library of templates (squares) and galaxy colors (grey circles) for MACS0416. As seen in the image, it was necessary to include a new redder template (EL1) to the original library for the reddest galaxies in the clusters.}
\label{redtemplate}
\end{center}
\end{figure}

While characterizing the quality of our photo-z estimates, we noticed that several cluster galaxies were retrieving unexpectedly poor photo-z estimates (when compared to others of similar magnitudes). In order to understand the origin of such dispersion, we ran \texttt{BPZ2.0} again on the whole spectroscopic sample (Section \ref{GCphotoz}) but using this time the ``ONLY\_TYPE=yes" mode to redshift the entire library of templates to the corresponding redshift and be able to compare the predicted and the observed colors according to the most likely template. When representing the color difference between models and data as a function of the (previously estimated) photo-z error, we found an (almost) linear relation between the two. This trend was indicating that several galaxies with peculiar colors (outside to color-space covered by the \texttt{BPZ2.0} templates, see Figure \ref{redtemplate}) were retrieving unexpectedly poor photo-z. It is worth noting that since \texttt{BPZ2.0} has to converge to a solution, the only way for the code to compensate such peculiar colors was to artificially shift the library of templates up and down in redshift up to find a solution minimizing the differences between data and models. This effect was causing the photo-z estimates to show a rather large scatter around the cluster redshift.  

\vspace{0.2cm}

To solve this issue, we used the template set of spectral energy distributions (SEDs) of luminous red galaxies (LRGs) from \citep{2013ApJ...768..117G}\footnote{These models, specifically selected to match the Sloan Digital Sky Survey (SDSS; \citealt{2000AJ....120.1579Y}) colors at different redshift bins, are generated by superposing model SEDs of composite stellar populations (CSP) with a burst model, allowing both components to be reddened by dust.} to identify a potential new template capable to reproduce the observed colors of these very red galaxies. In every cluster, we identified all cluster galaxies outside the \texttt{BPZ2.0} template color-space and estimated the mean color of such population. Then we redshifted the library of LRG models to the cluster redshift, computed their expected colors and kept the three templates providing closer matches. After repeating the same exercise for the 25 clusters, we ended up selecting the most favored template (\texttt{sedfit\_restframe\_z02\_507.sed}) and incorporating it to our library. After rerunning \texttt{BPZ2.0} we noticed that the new template was indeed improving the SED-fitting for such galaxies (retrieving a lower $\chi^{2}$ value) and broadening the spectral-type distribution of red galaxies in clusters and that the photo-z error was considerably reduced. An in-depth analysis of the physical properties of these galaxies will be addressed in a separate paper, investigating the possibility of being dusty star-forming galaxies \citep{2005A&A...443..435W}; a particular type of SED not included in most libraries of galaxy models.

\subsection{Performance on Cluster Members}
\label{GCphotoz}

In order to be able to characterize the final precision achieved for our photo-z estimates, it was necessary to compile a sample of galaxies with spectroscopic redshifts. The CLASH survey has been awarded with 225h of time on the Very Large Telescope (VLT) as a part of a ESO Large Programme (CLASH-VLT; PI: Piero Rosati) using the Visible Multi-Object Spectrograph (VIMOS; \citealt{2003SPIE.4841.1670L}) to obtain spectroscopic follow-up for 2000-4000 galaxies (as much cluster members as faint lensed galaxies) for each of the 13 southern clusters over a 20-25 arcmin FoV \citep{2014Msngr.158...48R}. Spectroscopic targets were selected down to R = 24 AB magnitude, with a color selection based on two or three colors, which however is wide enough to include the full range of galaxy. The success rate in measuring reliable redshifts is typically around 75\%, averaged over all magnitudes (R$<$24). The efficiency in recovering galaxy members varies from 50\% in the core to $\lesssim$ 10\% in the cluster outskirts (at approximately two virial radii), and also depends on cluster richness. We refer the reader to \cite{2016ApJS..224...33B} for a further explanation about the spectroscopic target selection. It is worth noting that although one of the four VIMOS pointing was constantly locked on the cluster cores, allowing long exposures on the lensed galaxies (between 30 minutes and 4 hours), due to the complexity of allocating so many slits inside the WFC3/FoV ($\sim$1'), only a few hundredth objects in a limited number of clusters were available within the innermost part of the CLASH clusters.   

\vspace{0.2cm}

\begin{figure}
\begin{center}
\includegraphics[width=8.cm]{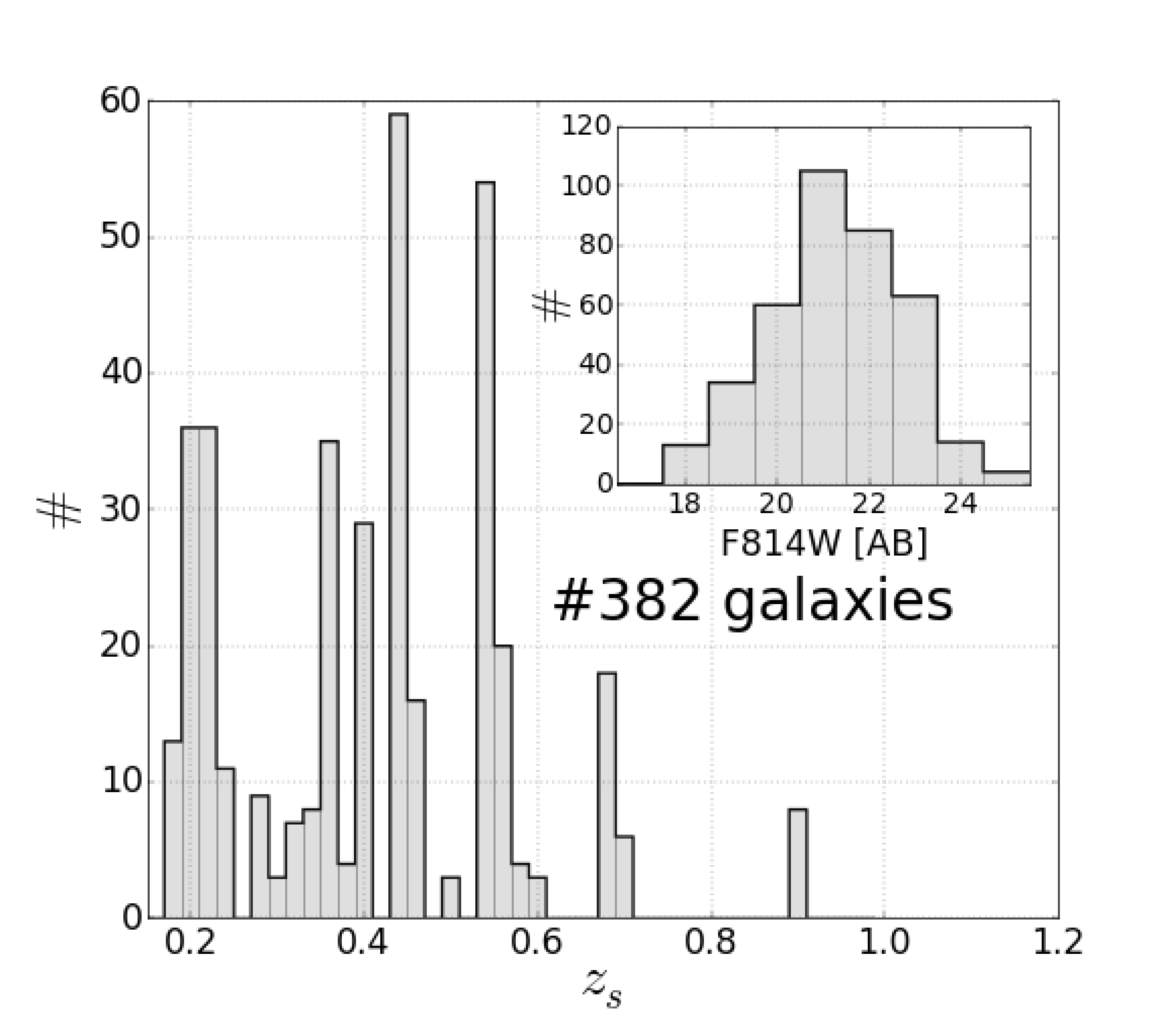}
\caption{In order to characterize the photo-z precision of our photometry, we compiled a sample of 382 galaxies spectroscopically confirmed to be cluster members. This control sample covers the entire $0.1<z<0.9$ redshift range of the CLASH clusters (main panel) and a range of $17<F814W<25$ in magnitude (inset panel).}
\label{spzsample}
\end{center}
\end{figure}

Along with the VLT data, in this paper we also collected spectroscopic redshift measurements from the Grism Lens-Amplified Survey from Space (GLASH; \citealt{2014ApJ...782L..36S}; \citealt{2015ApJ...812..114T}) and NASA/IPAD Extragalactic Database (NED). A final sample of 382 spectroscopically confirmed galaxies within the WFC3/FoV over the 25 CLASH clusters was selected to estimate the performance of the CLASH photo-z estimations. This selection was made imposing two criteria: 1. the galaxies had to be detected at least on 14 bands (out of 16) to guarantee a good sampling of their SED and 2. the differences between the galaxy and the cluster redshift had to be smaller than (or equal to) 0.01 (i.e., $|z_{g}$-$z_{cl}|\le0.01$). As illustrated in Figure \ref{spzsample}, this control sample covers the entire $0.1<z<0.9$ redshift range of the CLASH clusters ($<z>=0.41$) and a range in magnitude of $17<F814W<25$ ($<m>=21.3$). 

\vspace{0.2cm}

\begin{figure*}
\begin{center}
\includegraphics[width=8.5cm]{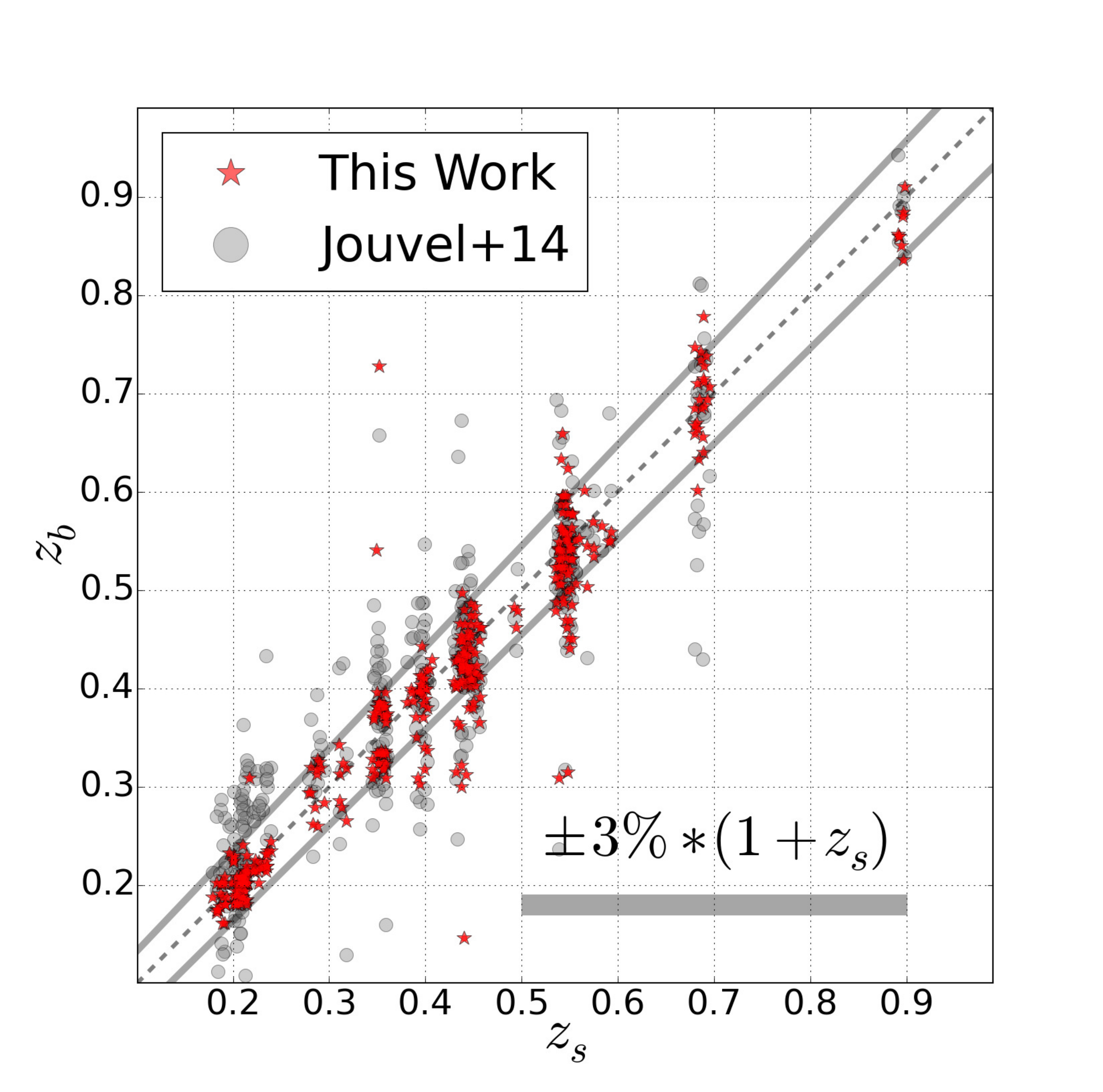}
\includegraphics[width=8.5cm]{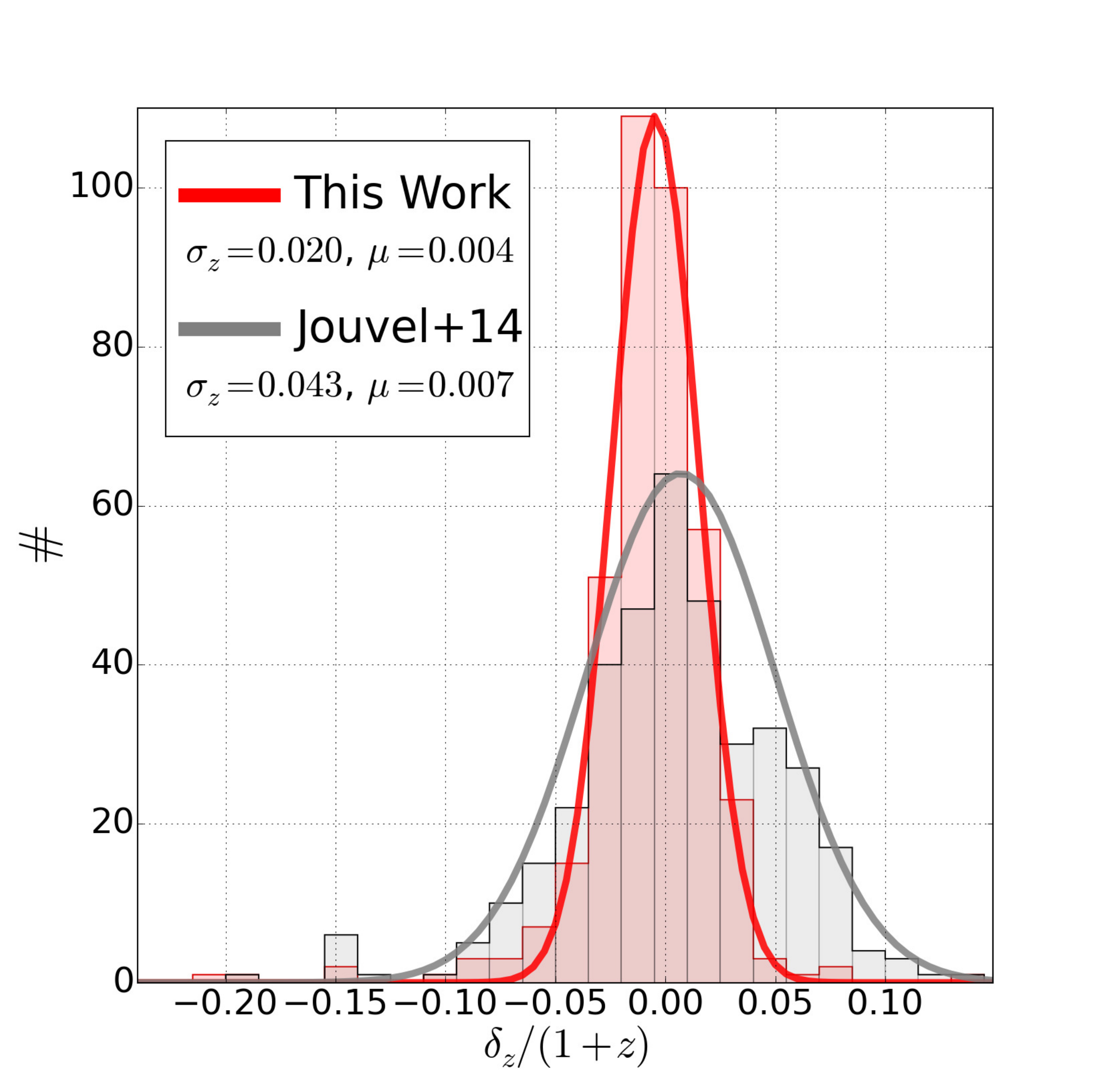}
\caption{The CLASH Photometric Redshift Accuracy for Cluster Members (I). Left: The figure compares the photometric redshift performance, as a function of the real spectroscopic redshift values, derived in this work (red stars) with those from our previous pipeline (J14, grey circles). This new photometry provides more accurate estimates reducing the typical observed scatter of cluster galaxies around the cluster redshift. Right: The figure compares the photo-z error distribution function derived in this work (red) with that from our previous pipeline (grey). This new photometry yields an overall precision of $\sigma_{z}$=0.020 (compared to the previous $\sigma_{z}$=0.042), representing a factor two improvement.}
\label{clashaccCMs1}
\end{center}
\end{figure*}

\begin{figure*}
\begin{center}
\includegraphics[width=8.5cm]{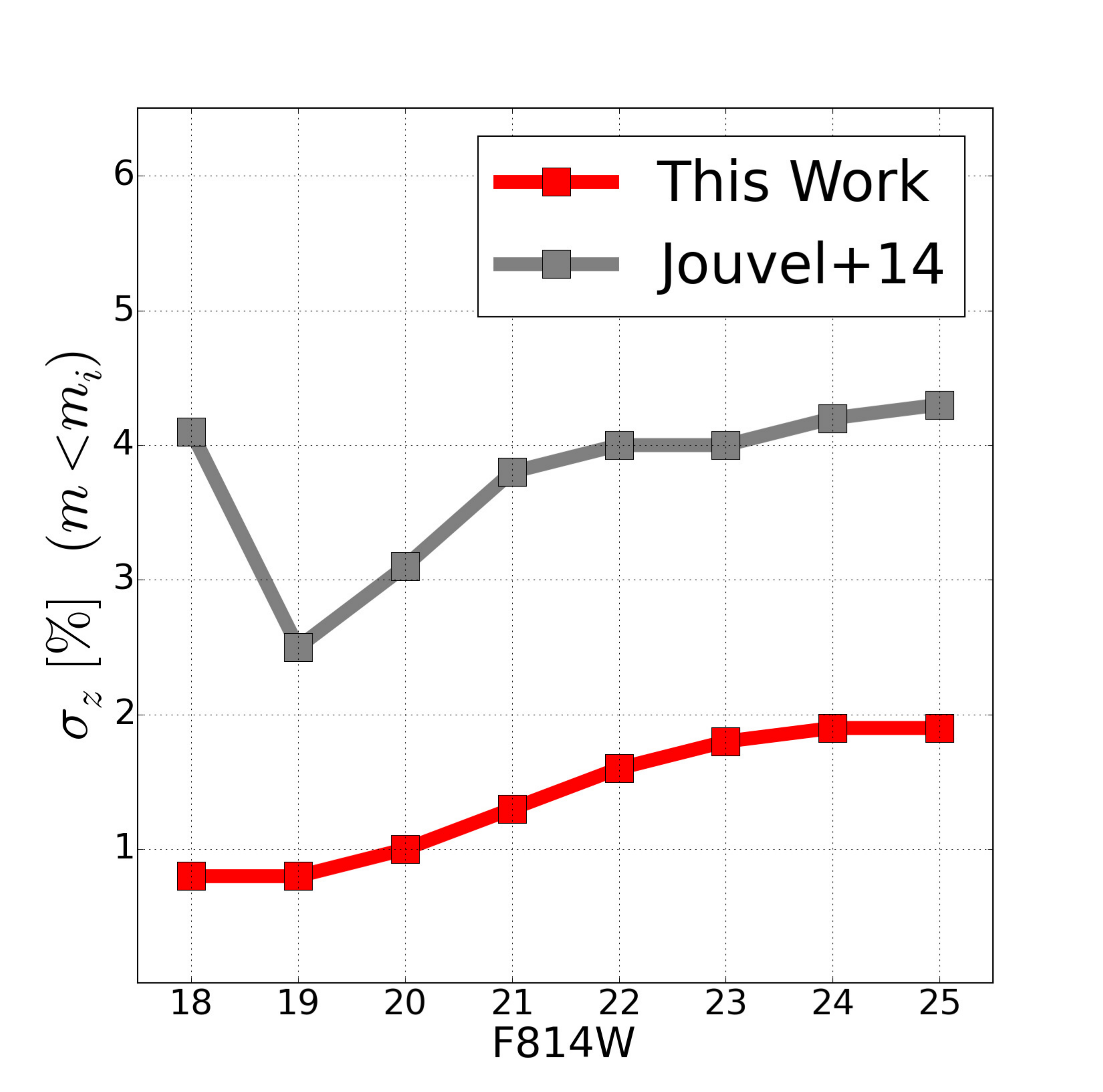}
\includegraphics[width=8.5cm]{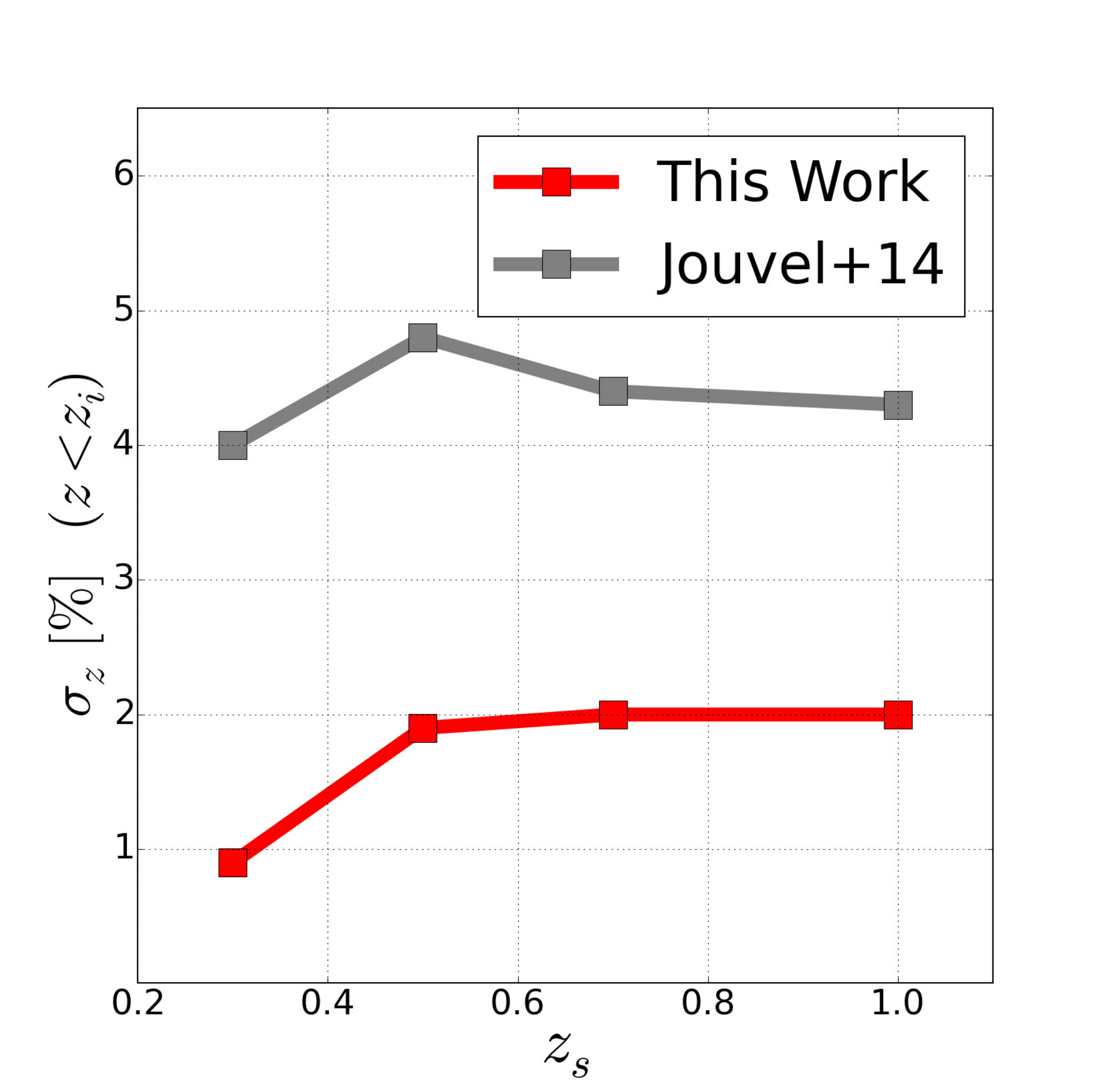}
\caption{The CLASH Photometric Redshift Accuracy for Cluster Members (II). Left: The figure compares the cumulative photometric redshift accuracy for Cluster Members ($\sigma_{z}$), as a function of an apparent magnitude (F814W), achieved in this work (red) with that from our previous pipeline (grey). This new photometry not only provides a higher overall precision at all magnitudes (an overall factor of 2 or up to a factor of 4 at F814W=18) but also (an expected) meaningful distribution of values as a function of the signal-to-noise (magnitude). Right: Similarly, the figure compares the cumulative photometric redshift accuracy for Cluster Members ($\sigma_{z}$), as a function of the redshift, achieved in this work (red) with that from our previous pipeline (grey). Again, this new photometry not only provides a higher precision at all redshift bins (an overall factor of 2 or up to a factor of 4 at z$<$0.3) but also a meaningful distribution of values as the galaxies become statistically fainter at higher redshift.}
\label{clashaccCMs2}
\end{center}
\end{figure*}

Based on this sample, our new photo-z estimates reach an accuracy of dz/1+z $\sim$0.8\% for galaxies brighter than magnitude F814W$<$18,  a dz/1+z $\sim$1.0\% for galaxies brighter than magnitude F814W$<$20, a dz/1+z $\sim$1.6\% for galaxies brighter than magnitude F814W$<$22 and a dz/1+z $\sim$2.0\% for galaxies brighter than magnitude F814W$<$23. Globally, this sample yields an accuracy of dz/1+z $\sim$2.0\% with an averaged magnitude $<F814W>$=21.3. The fraction of catastrophic outliers is always below 3\% except for the faintest magnitude bin ($23.5<m<24.5$) where the signal-to-noise of galaxies makes the photo-z estimation more uncertain. In terms of the redshift, the sample reaches an accuracy of dz/1+z $\sim$1.0\% for galaxies at redshifts $0.1<z<0.3$, of dz/1+z $\sim$2.2\% for galaxies at redshifts $0.3<z<0.5$ and of dz/1+z $\sim$2.4\% for galaxies at redshifts $0.5<z<0.7$. These results are illustrated in Figure $\ref{clashaccCMs1}$ \& $\ref{clashaccCMs2}$ and summarized in table $\ref{phzacctable}$. When this precision is compared to that presented in J14, we find (almost) a factor of two improvement at all magnitudes. As shown in Figure $\ref{clashaccCMs1}$, this improvement can be as high as a factor of three for high signal-to-noise galaxies (F814W$<$20).  

\vspace{0.2cm}

\begin{figure}
\begin{center}
\includegraphics[width=8.5cm]{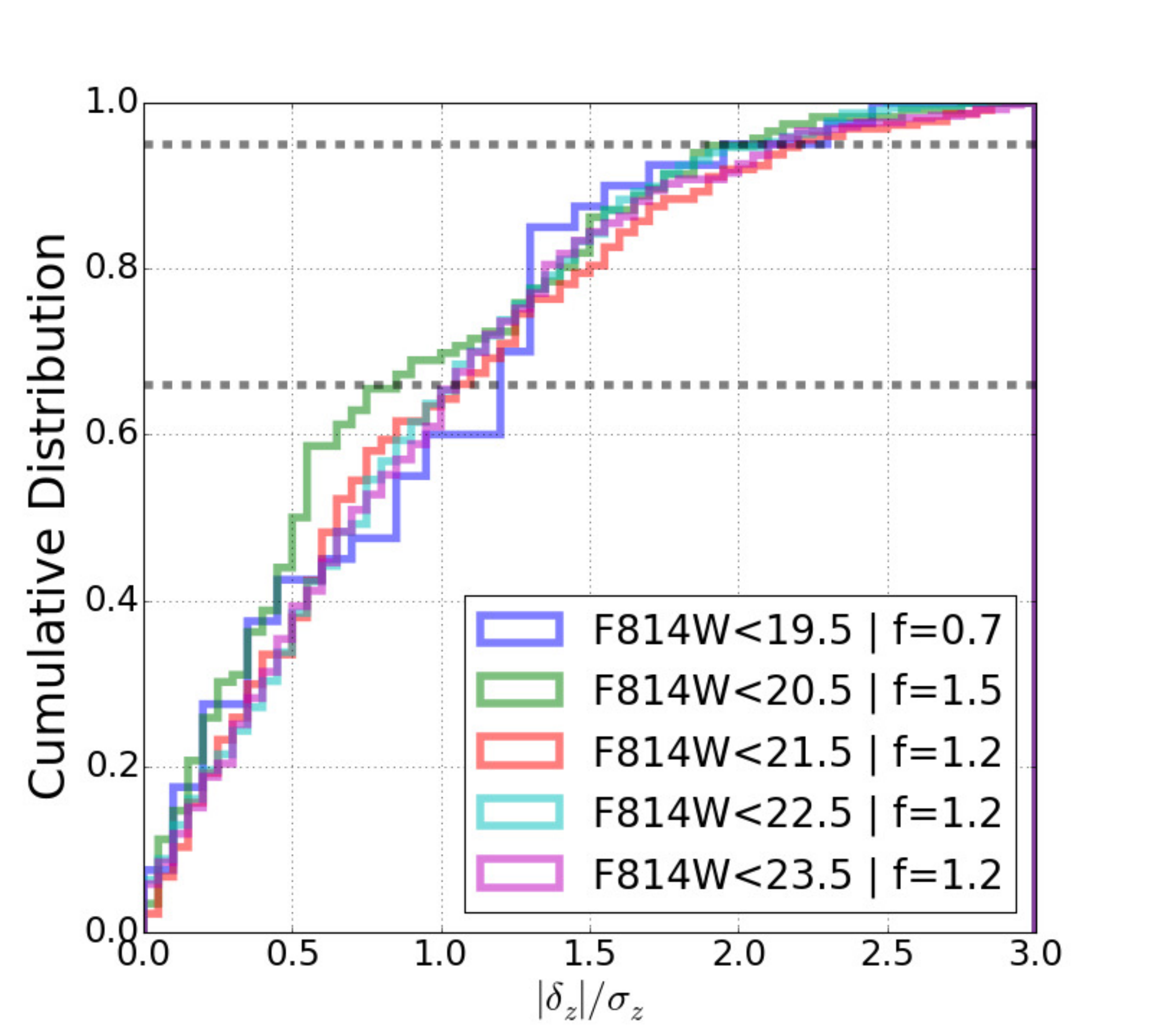}
\includegraphics[width=8.5cm]{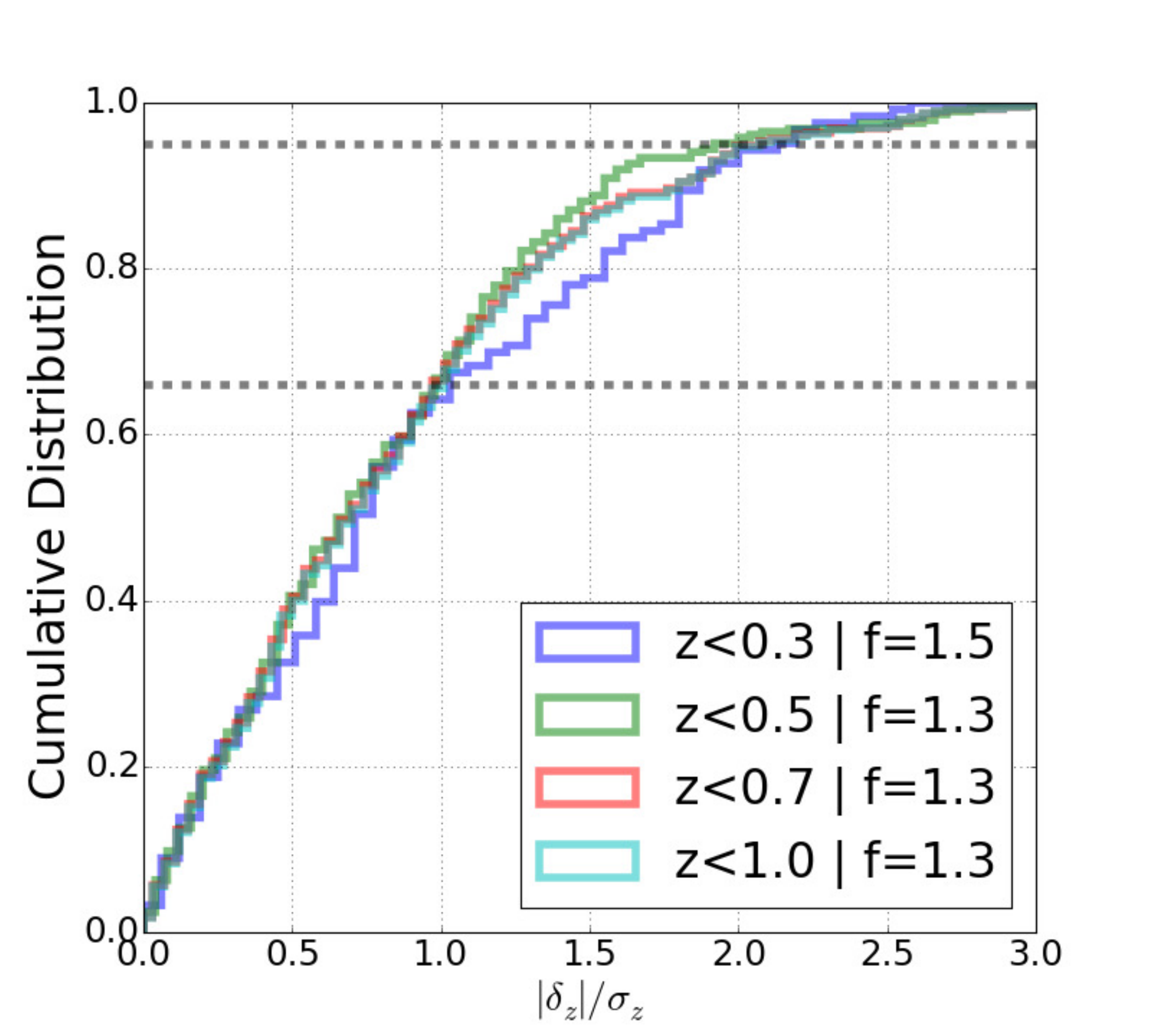}
\caption{The CLASH Photometric Redshift Accuracy for Cluster Members (III). The figure shows the cumulative photo-z error distribution function for six different magnitude (upper) and four redshift (lower) bins. Horizontal dashed grey lines correspond to the $\sim$64\% and $\sim$90\% within the formal $1\sigma$ and $2\sigma$ confidence interval, respectively. An additional multiplicative factor (\textit{f}) is needed for the distributions to describe a Gaussian function. These corrections may indicate that the photo-z precision may either be a bit underestimated ($<0.5\%$) at bright magnitudes and a bit overestimated ($>0.5\%$) at faint magnitudes.}
\label{clashaccCMs3}
\end{center}
\end{figure}

In order to verify that dz/1+z is representative for the spectroscopic sample, the cumulative distribution of dz/1+z is represented in Figure\ref{clashaccCMs3}. We observed that an additional multiplicative factor (\textit{f}) needs to be applied to our estimates to retrieve the $\sim$64\% and $\sim$90\% of the photometric redshifts within the formal $1\sigma$ and $2\sigma$ confidence interval, respectively. The corrections, shown in Figure \ref{clashaccCMs3}, indicate that the photo-z accuracy is a bit underestimated ($<$0.5\%) at bright magnitudes and a bit overestimated ($>$0.5\%) at faint magnitudes. Table $\ref{phzacctable}$ includes the expected accuracy for the CLASH photo-z with and without applying this additional factor. 

\begin{table}
\caption{Comparison of the photometric redshift performance for a spectroscopically confirmed sample of 428 cluster as a function of both apparent magnitude in the F814W and redshift. Original result from J14 are marked as $\sigma_{z,o}$.}
\begin{center}
\label{phzacctable}
\begin{tabular}{|l|c|c|c|c|c|c|c|c|c|c|c|c|c|c|}
\hline
\hline
\,\,\,\,\,\,\,\,\,\,\,\,\, F814W   & $\sigma_{z}$ & $\sigma_{z,o}$ & $\mu_{z}$ & $\#$ & $\eta$ [$\%$] \\
\hline
18.5 $<$ m $<$ 19.5 & 0.008 & 0.036 & 0.004 & 34 & 0.0  \\
19.5 $<$ m $<$ 20.5 & 0.013 & 0.032 & -0.001 & 60 & 0.0  \\
20.5 $<$ m $<$ 21.5 & 0.016 & 0.030 & 0.000 & 105 & 0.9  \\
21.5 $<$ m $<$ 22.5 & 0.020 & 0.044 & 0.008 & 86  & 1.2 \\
22.5 $<$ m $<$ 23.5 & 0.022 & 0.051 & 0.013 & 63 & 3.2  \\
\hline
\,\,\,\,\,\,\,\,\,\,\,\,\,\,\,\,\, $z_{sp}$   & $\sigma_{z}$ & $\sigma_{z,o}$ & $\mu_{z}$ &  $\#$ & $\eta$ [$\%$] \\
\hline
0.10 $<$ z $<$ 0.30 & 0.009 & 0.040 & 0.005 &  108 & 0.0  \\
0.30 $<$ z $<$ 0.50 & 0.022 & 0.046 & 0.007 &  161 & 2.5 \\
0.50 $<$ z $<$ 0.70 & 0.024 & 0.029 & 0.005 &  105 & 2.9  \\
0.70 $<$ z $<$ 1.00 & 0.011 & 0.020 & 0.014 &     8  & 0.0 \\
\hline
\hline
\end{tabular}
\end{center}
\end{table}

\subsection{Photometric zero-point calibrations.}
\label{pzp}

It is now customary for most groups deriving photo-z estimations to end up performing some sort of photometric zero-point corrections on the input photometry.  Although they neatly improve the final photo-z performance (in terms of both accuracy and fraction of outliers), the provenance of these corrections is most times uncertain (see \citealt{2014MNRAS.441.2891M} for an in-depth discussion). Main explanations range from systematic differences among colors of stars and galaxies (to compensate unnoticed biases while performing the multi-band photometry), systematic issues during the data reduction process or as a consequence of faulty calibrations of galaxy models since different libraries typically yield slightly different corrections.

\vspace{0.2cm}

When using template-based photo-z codes, it turns out possible to compare the expected and the observed colors (fluxes) for the galaxies. If the photometry is assumed to be accurate, the library of models reliable (in terms of both calibration and completeness) and the galaxy redshifts are known (so the templates can be redshifted to those values), the observed scatter between expected and measured colors is supposed to be solely caused by the inherent photometric noise in images. Since this background signal can be approximated as a normal distribution with null mean ($\mu=0$), the dispersion between colors (i.e., the ratio among fluxes) is therefore expected to be also a normal distribution with mean equal 1 ($\mu=1$) and a dispersion proportional to this background noise. If that assumption applies, any statistical deviations ($\mu \neq 1$) are assumed to be an instrumental zero-point offset to be corrected.

\vspace{0.2cm}

In general, empirical galaxy templates are typically calibrated (and tested) using large samples of galaxies at different redshift and magnitude ranges. That is why these models provide accurate photo-z estimations on average. However, considering the reduced sample of galaxies with spectroscopic redshifts in the cores of our 25 clusters along with their particular redshift distribution (galaxies clustered at specific ranges), it turns out risky to try to systematically re-calibrate the photometric zero-points for each cluster individually. There is no guarantee that the templates will faithfully reproduce the colors of galaxies at a very specific redshift (see section \ref{GCphotoz}). In those situations, the so-derived corrections may certainly improve the photo-z precision for a subsample of galaxies but they could artificially lead to biases for other galaxies at different redshifts.  

\begin{figure}
\begin{center}
\includegraphics[width=8.5cm]{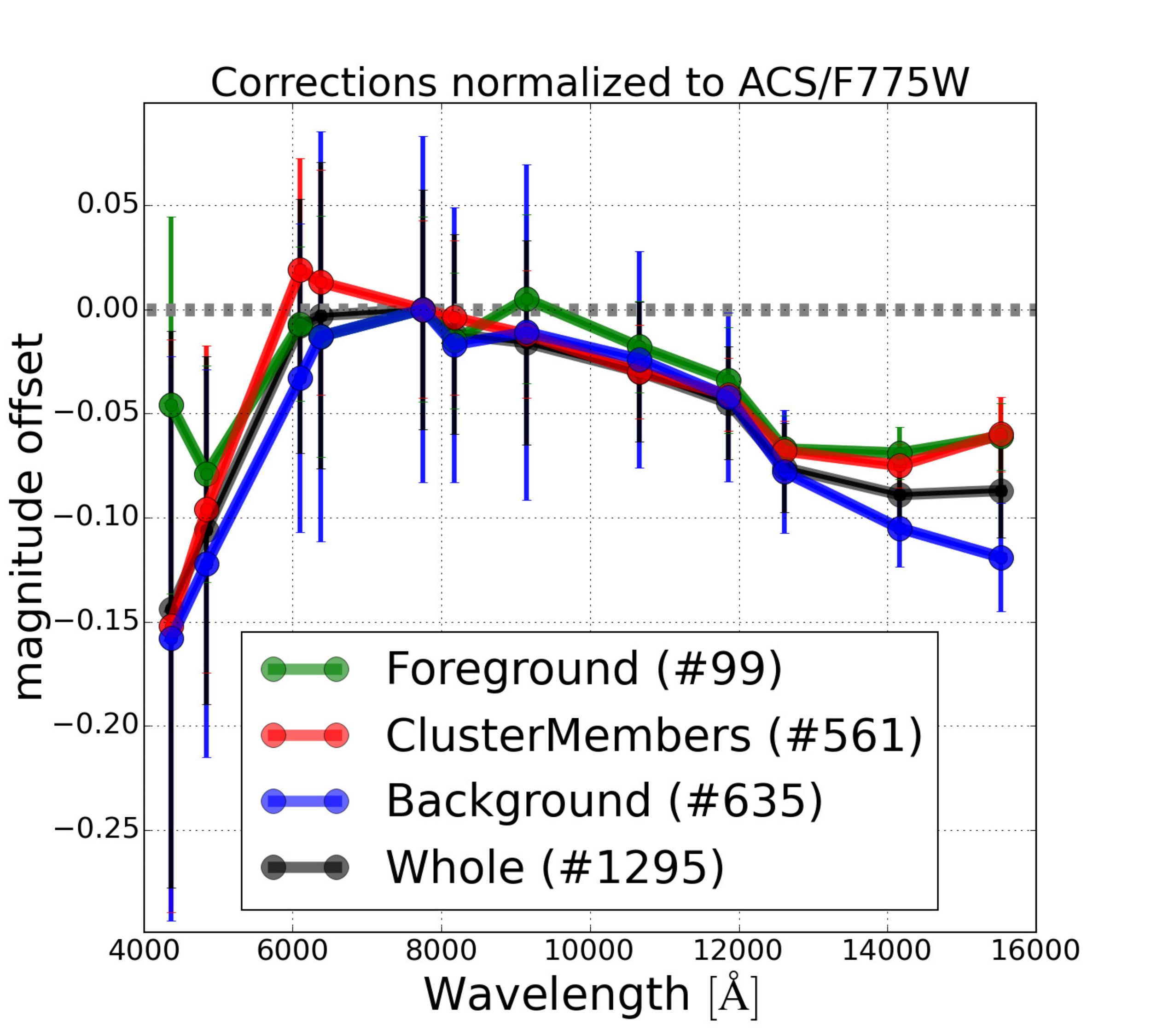}
\caption{Magnitude corrections for the HST data (I). The figure shows the magnitude corrections derived as a function of wavelength from a sample of 1295 galaxies with known spectroscopic redshifts. The sample is divided in four groups: foreground (green), cluster (red), background (blue) and all (black) galaxies. A trend is observed for all types of galaxies indicating that NIR magnitudes are systematically fainter than those expected by the \texttt{BPZ} models.}
\label{zpoff1}
\end{center}
\end{figure}

\begin{figure}
\begin{center}
\includegraphics[width=8.5cm]{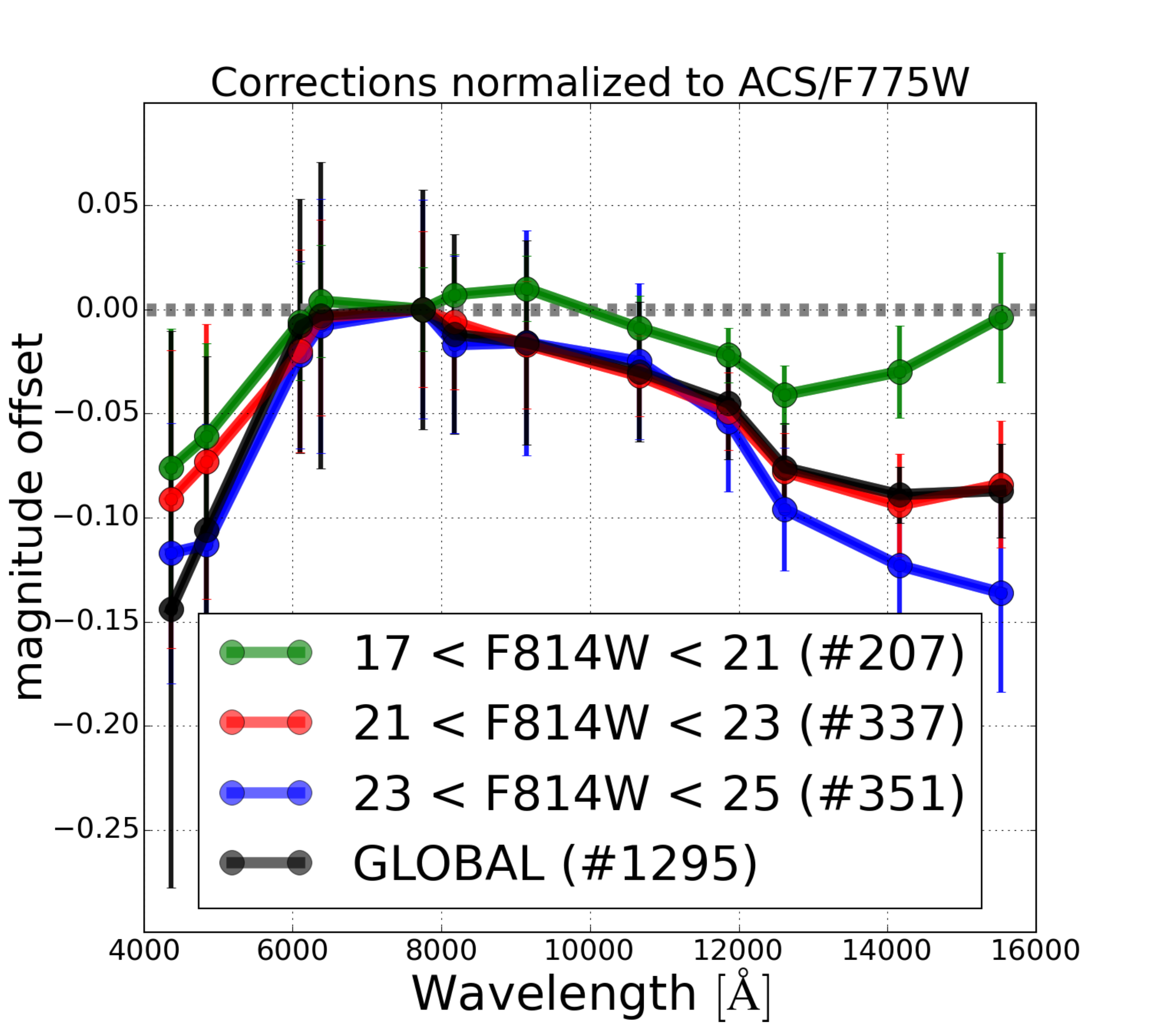}
\caption{Magnitude corrections for the HST data (II). The same as in Figure \ref{zpoff1} but dividing the sample in three magnitude bins: $17<$F814W$<21$,  (green), $21<$F814W$<23$ (red) and $23<$F814W$<25$ (blue). As before, it is observed a systematic trend with wavelength showing a clear spread in magnitude where fainter galaxies depart more from models.}
\label{zpoff2}
\end{center}
\end{figure}

\begin{figure}
\begin{center}
\includegraphics[width=8.5cm]{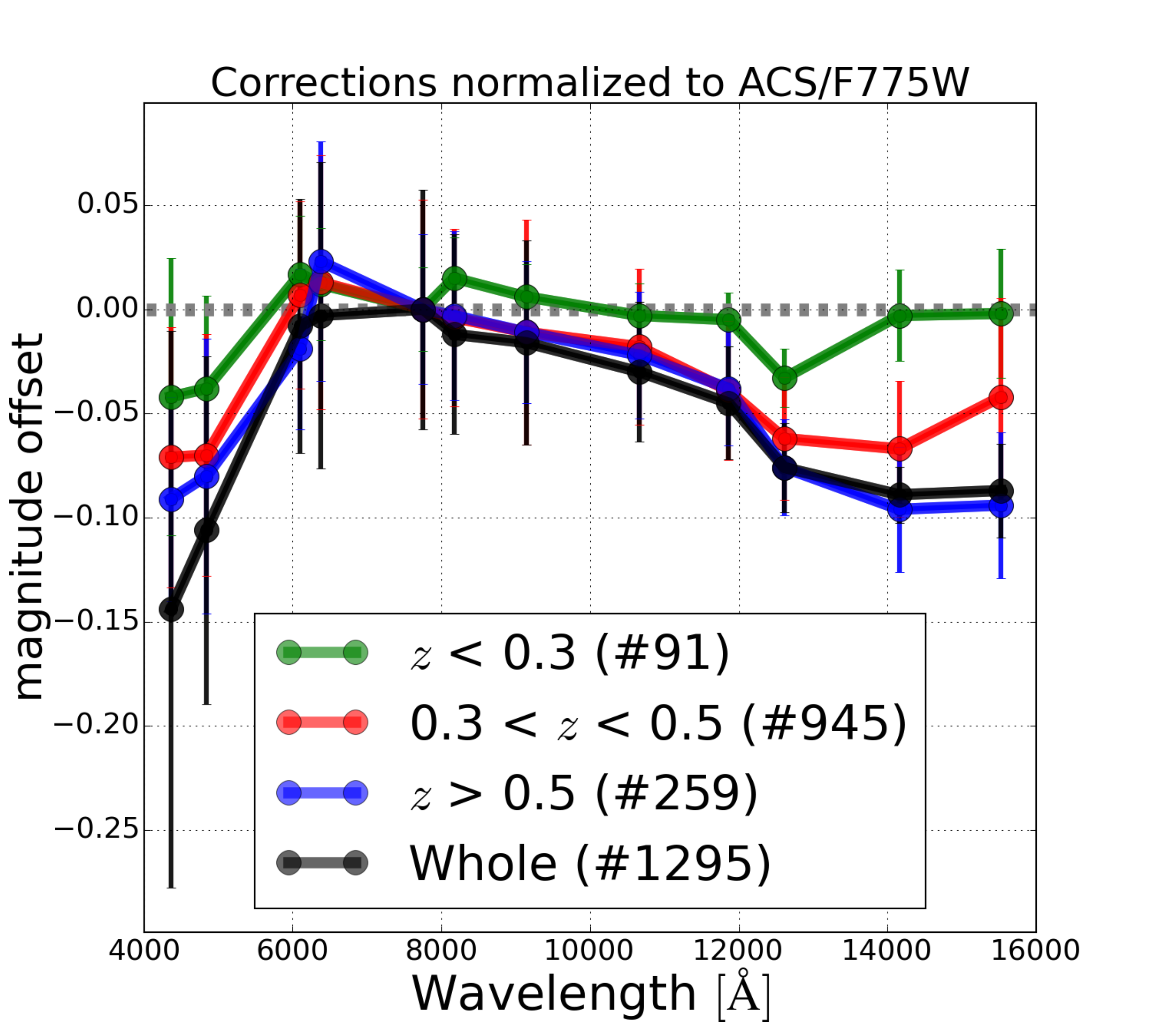}
\caption{Magnitude corrections for the HST data (III). The same as in Figure \ref{zpoff1} \&  \ref{zpoff2} but dividing the sample in three redshift bins: z$<0.3$ (green), $<0.3<z<0.5$ (red) and z$>0.5$ (blue). Once again, it is observed a systematic trend with wavelength where high-z galaxies show a large deviation from expectations.}
\label{zpoff3}
\end{center}
\end{figure}

\begin{figure}
\begin{center}
\includegraphics[width=8.5cm]{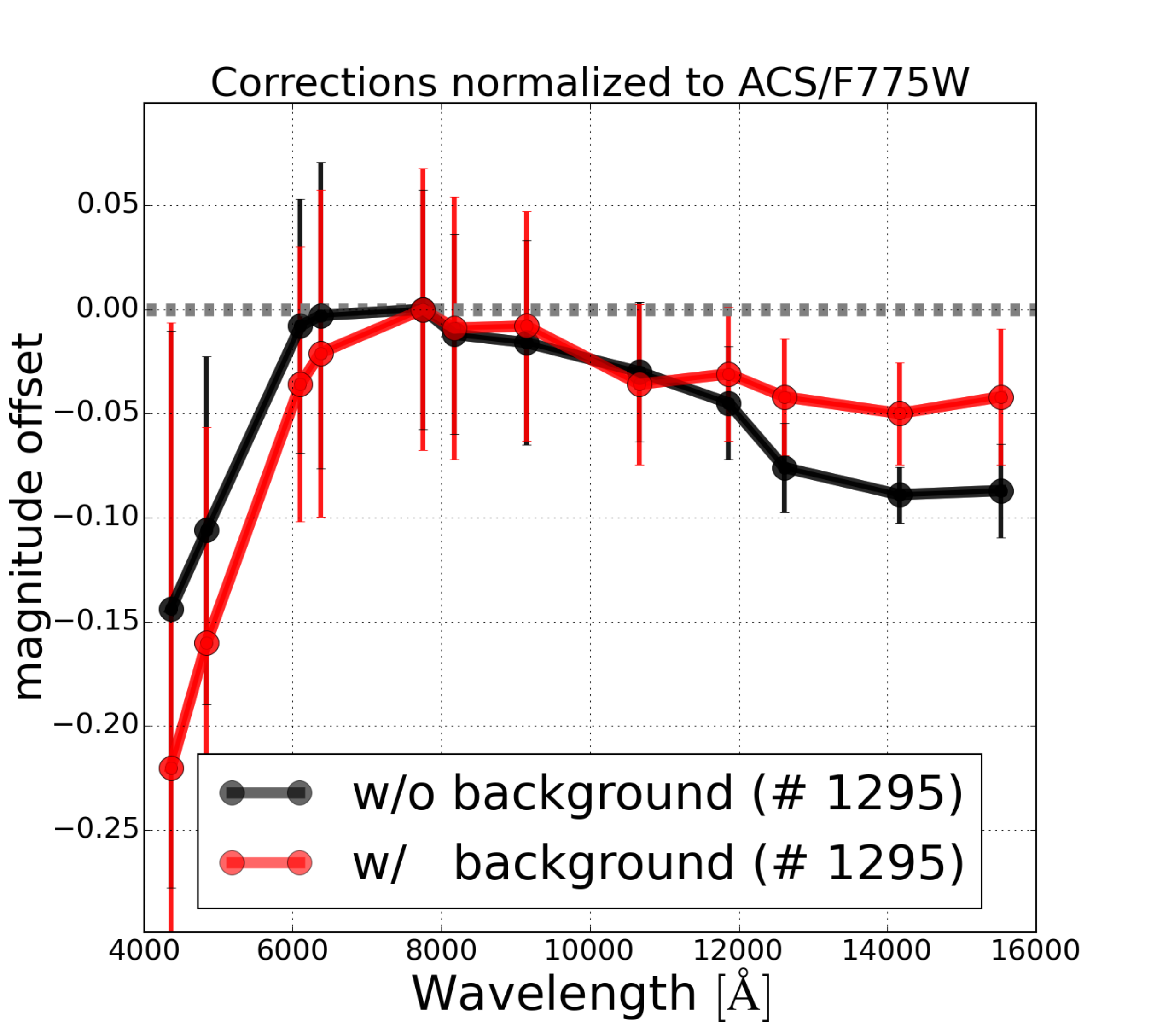}
\caption{Magnitude corrections for the HST data (IV). The same as in Figure \ref{zpoff1}, \ref{zpoff2} \&  \ref{zpoff3} but showing the magnitude corrections derived after (black) and before (red) removing the BCL signal from images. Once again, it is observed a systematic trend with wavelength. Whereas the differences between data and models in the NIR were slightly reduced in this case, the magnitudes in the optical were showing the opposite behavior.} 
\label{zpoff4}
\end{center}
\end{figure}

\vspace{0.2cm}

The photometric zero-point of an instrument is, by definition, the magnitude of an object that produces 1 count per second \citep{2005PASP..117.1049S}. These instrumental quantities are therefore independent of the observed targets. If there was a real zero-point offset in a given dataset, it would indicate that different galaxy populations might agree on the  corrections to be applied. In order to explore the need of zero-point offset for our photometry, we followed a similar approach as that presented in \citep{2014MNRAS.441.2891M} where the spectroscopic-z sample is divided in several categories according to their colors, redshifts and magnitudes. Whereas the first two analysis might flag issues with the library of models, the last one might indicate issues with the photometry.
 
\vspace{0.2cm}

Initially, we separated the whole sample among foreground, cluster and background galaxies and re-estimated offsets per each subsample. Since these populations mainly correspond to different galaxy templates, finding similar offsets may indicate that the observed bias is not caused by the library of models. It is worth recalling that the three categories will include galaxies at different redshift ranges, making the corrections less sensitive to particular template calibration issues at specific redshifts\footnote{The \texttt{BPZ} library of templates was calibrated using optical HST data.}. As seen in Figure $\ref{zpoff1}$, we observe a clear trend in the computed offsets as a function of the wavelength, from the optical to the NIR, indicating that galaxies are systematically fainter than expected by the models. Whereas this effect is barely observed in the optical range ($6000<\lambda<9000\AA$), it becomes clearer in the NIR ($\lambda>10000\AA$). Although background galaxies (blue line) deviate more than other types in the NIR, the three categories show the same pattern. Therefore, we think this effect might not be assigned to the models. 

\vspace{0.2cm}

Later on, we separated the sample in three similar-size magnitude bins: $17<$F814W$<21$, $21<$F814W$<23$ \& $23<$F814W$<25$ and re-estimated the offsets. As before, each subsample may include galaxies with different colors and redshifts, making the offsets less sensitive to the library of models. In this case, similar corrections might point to the photometry as the main reason for the observed bias. As seen in Figure $\ref{zpoff2}$, the average (black) line shows the same trend although it is observed a clear spread in magnitude. Whereas for bright galaxies the deviation is notably reduced at all wavelengths, the effect is dramatically increased for the faintest galaxies.

\vspace{0.2cm}   

Finally, we separated the sample in different redshift bins: $z<0.3$, $0.3<z<0.5$, $z>0.5$ and re-estimated the offsets. As in the former cases, although a certain color-magnitude evolution is expected for the galaxies as a function of redshift, each subsample may include galaxies at different magnitudes and colors (cluster members and background galaxies), making the corrections less sensitive to the models. As seen in Figure $\ref{zpoff3}$, we observe a similar result as that obtained when segregating the galaxies according to their apparent magnitudes. Again, we observe the same wavelength-dependent pattern in the data, where low-z galaxies show a moderate bias at all wavelengths while high-z galaxies seem to deviate more from the expectations. This result is consistent with that obtained when dividing galaxies per magnitude bins, since low-z galaxies will be (in average) brighter than galaxies at high-z.   

\vspace{0.2cm}

As a complementary analysis, we investigated if the background subtraction from images might be responsible for those offsets. Initially, we restored the background-signal removed by \texttt{SExtractor} from the galaxies and re-computed the offsets. Unfortunately, this additional signal was incapable to alleviate the tension between data and models. Secondly, we re-ran \texttt{SExtractor} without performing any background subtraction at all from the images. As illustrated in Figure \ref{zpoff4}, although we observed in this case that the differences between data and models in the NIR were slightly reduced, the magnitudes in the optical were showing the opposite behavior; i.e., increasing the tension between data and models. 

\vspace{0.2cm}

In the light of what has been presented above, there seems to be several explanations. Perhaps the simplest one would be due to inaccuracies during the PSF homogenization of images. We observe that the offsets get worse at shorter and longer wavelengths respect to the ACS/F775W band which happens to have a sharper PSF than the WFC3. Another interpretation may be a tension between the \texttt{BPZ} models and data since the former does not include specifically any physical evolution and the bias increases with redshift. Finally, since the bias increases as decreases the signal-to-noise of galaxies, there could be an instrumental issue with the WFC3 camera where, at a faint count regime and long wavelengths, the detector departs from linearity not integrating as much photons as expected. Similar systematics have been reported from other modern astronomical imagers (\cite{2014JInst...9C3048A}, \cite{2015A&A...575A..41G}). This scenario will be explored in a separate paper using the data from the CANDELS (\cite{2011ApJS..197...35G}; \cite{2011ApJS..197...36K}) fields.

\vspace{0.2cm}

\begin{figure}
\begin{center}
\includegraphics[width=8.5cm]{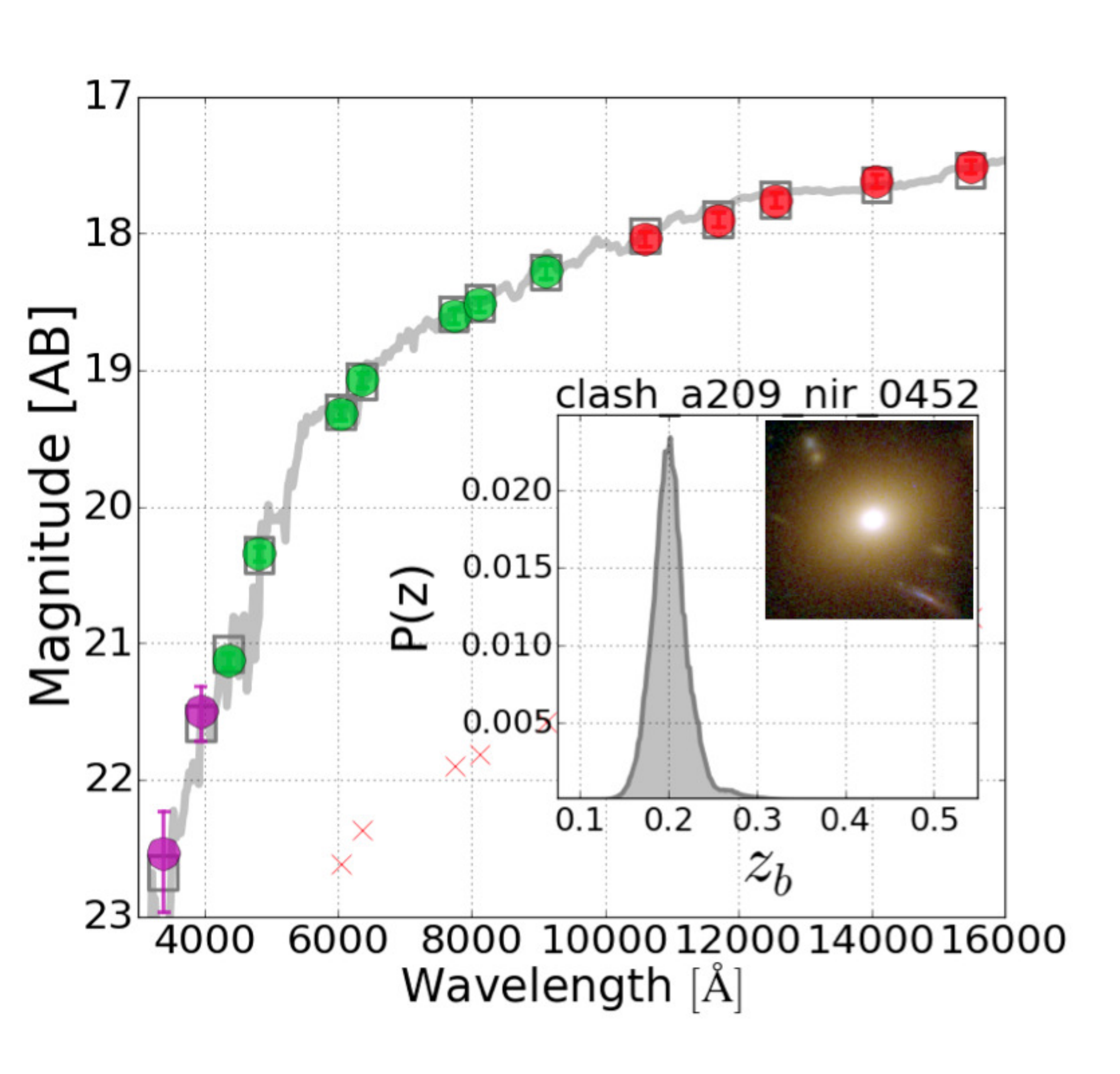}
\caption{The figure shows an example of the obtained SED-fitting for an early-type cluster galaxy with the new photometry derived in this work, where different wavelength ranges are represented with different colors (UV/purple, Optical/green \& NIR/red). Inset panel shows the corresponding single-peak redshift probability distribution function P(z).}
\label{photozexamples}
\end{center}
\end{figure}

Irrespectively of the source of these offsets, it is important to emphasize that averaged zero-point corrections (using the whole spectroscopic sample at once) may not improve the overall quality of the CLASH photo-z, only benefiting a fraction of the galaxies. Meanwhile, the photometric redshift performance presented in Section \ref{GCphotoz} was estimated after applying these empirical magnitude-dependent offsets to the original photometry. It is worth noting that after implementing these corrections, the overall photo-z precision was dramatically improved. An example of the obtained SED-fitting for an early-type cluster galaxy with the new photometry is shown in Figure \ref{photozexamples}. Tables \ref{CLASHFS1}, \ref{CLASHFS2} \& \ref{CLASHFS3} summarize the observed magnitude offsets presented during this section.

\subsection{Photometric Uncertainties (II).}
\label{photoerr2}

While checking the quality of our photo-z estimations, we found an unexpectedly large fraction of catastrophic outliers (see Section \ref{BPZ}) at bright magnitudes. We noticed that several well-isolated early-type galaxies with high signal-to-noise and secure photometry (according to \texttt{SExtractor}) were getting completely wrong redshift estimates. After a careful inspection of the PDF of each galaxy, we noticed that usually the redshift distribution of these galaxies did not even include the correct redshift value; i.e., suggesting that the (real) redshift of the galaxy was incompatible with the observed colors. 

\vspace{0.2cm}

\begin{figure}
\begin{center}
\includegraphics[width=8.cm]{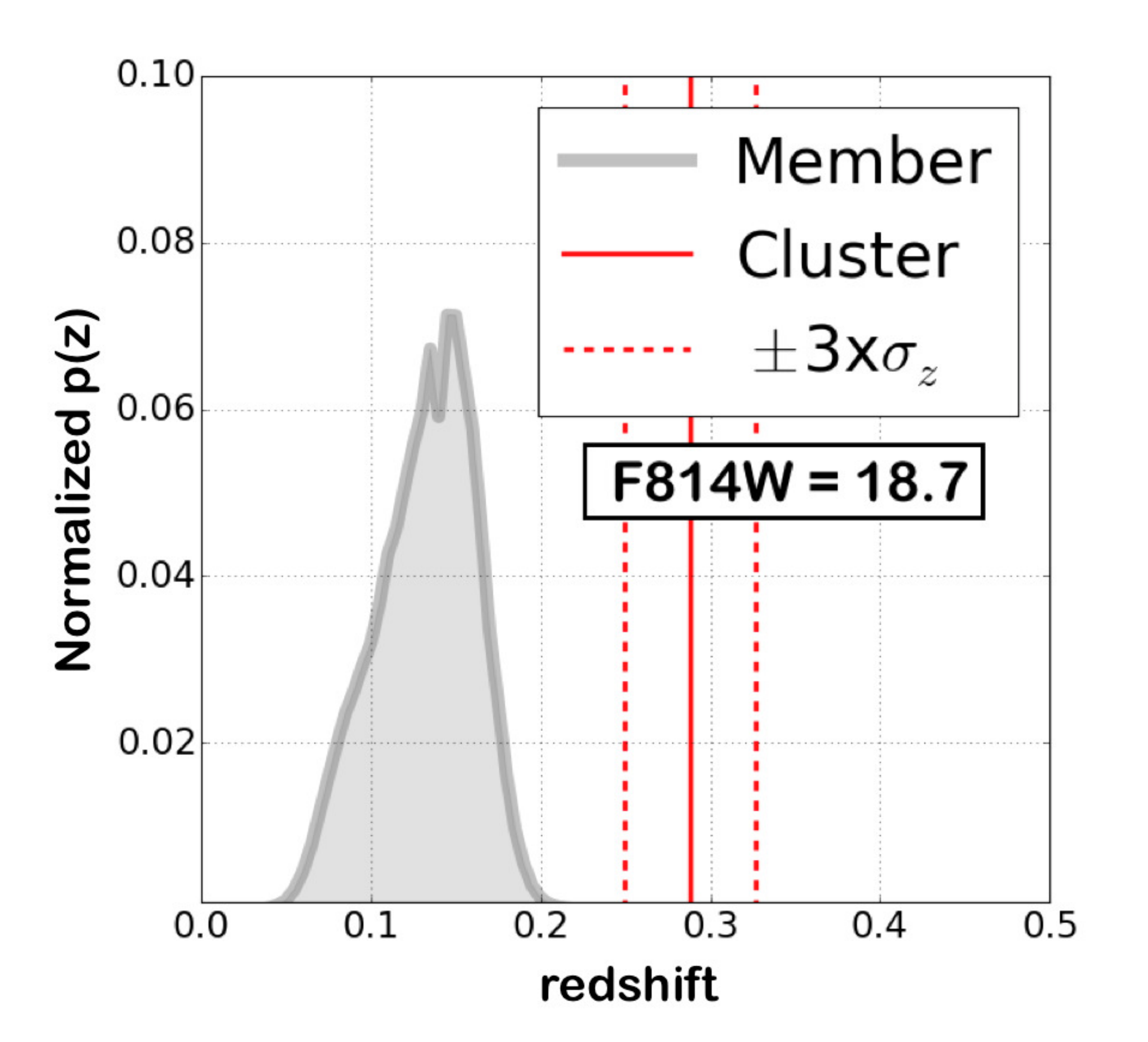}
\caption{Example of a faulty unimodal p(z) for a bright galaxy. Underestimated uncertainties were preventing the \texttt{BPZ} code to fully map out the z-T space, causing the likelihood to stack in a relative (nor necessarily an absolute) minimum. These biased single-peak distributions were artificially making the \texttt{Odds} parameter to be high (large confidence) despite of the wrong photo-z estimates. The problem was circumvented after adding an additional term in quadrature to the photometric uncertainties.}
\label{FailedPDF}
\end{center}
\end{figure}

As discussed in Section \ref{pzp}, when using photo-z codes based on model fitting, it becomes feasible to compare expected colors from models with real data if the redshift of the galaxies in known beforehand. When these differences are weighted by its corresponding photometric noise, the final error distribution is supposed to be well-approximated to a Gaussian function with mean equal to 1 irrespective of the magnitude range. This simple comparison encodes the level of agreement (or disagreement) between data, models and photometric uncertainties. We discovered that due to the fact that the photometric uncertainties were enormously underestimated for those bright sources, the \texttt{BPZ} code was unable to properly map out the entire redshift - spectral-type (z-T) space when computing the likelihood. Causing the analysis to stack in a relative (not necessarily an absolute) minimum. When projected in redshift-space (marginalizing over types), this effect was causing the resulting p(z) to be (generally) unimodal but placed at the wrong position. Importantly, this faulty unimodal p(z) was artificially making the \texttt{Odds} parameter from the \texttt{BPZ} to be high (meaning high confidence) but with a completely wrong redshift, making unfeasible to rely on this parameter to select accurate and reliable samples. This is the reason why in J14 was not possible to reliable isolate accurate photo-z estimates for bright cluster galaxies using the \texttt{Odds} parameter. This effect is illustrated in the Figure \ref{FailedPDF} where we show the PDF of a cluster galaxy and the 3$\times \sigma_{z}$ interval around the cluster redshift.  

\vspace{0.2cm}

In order to circumvent this problem, it was necessary to add an additional noise term in quadrature to the photometric uncertainties in every band to make the aforementioned distributions to compensate the differences between models and data. The noise terms were derived using all the spectroscopically confirmed cluster galaxies with magnitudes brighter than F814W$<$20. The exact nature of this excess noise (or uncertainty) is unclear but reflects the fact that the dominant error for these bright objects in no longer Poisson noise. It may be due to non-reported systematics when computing our photometry (such as imperfect PSF-homogenization) or a tension between data and models at very high signal-to-noise ratio (i.e., an imperfect library of galaxy models) as discussed in \cite{2006AJ....132..926C}. 

\section{Photometric Redshift Catalogue}
\label{catalogues}

We have run \texttt{BPZ} on this new CLASH photometry to generate a new photometric redshift catalogue for the 25 CLASH massive galaxy clusters. The catalogue, described in this section, includes both astrometric, morphologic, photometric and photo-z information for all detected sources in an NIR detection image (e.g., a weighted sum of WFC3/IR images: F105W, F110W, F125W, F140W, F160W).

\vspace{0.2cm}

Unique IDs are given to every detection according to the detection image and the cluster field from which its was detected following the next criteria: ID = clash\_clj1226\_nir\_0001 stands for \texttt{clash} (HST programme) + \texttt{clj1226} (galaxy cluster) +  \texttt{nir} (photometric aperture defined according to the NIR detection image) + \texttt{0001} (SExtractor\_ID). Both astrometric and geometrical information is therefore derived from its corresponding NIR detection image. The catalogue includes several parameters regarding both astrometric and morphological information from sources: celestial coordinates (\texttt{RA},\texttt{Dec}) in the J2000 system, physical position on the CCD (\texttt{X},\texttt{Y}), photometric aperture size (\texttt{AREA}), compactness (\texttt{FWHM}), the signal-to-noise (\texttt{s2n}) defined as $SExt\_FLUX\_AUTO/SExt\_FLUXERR\_AUTO$ on the detection image, standard SExtractor photometric Flags (\texttt{photoflag}), the number of filters a source was observed (\texttt{nfobs}) and detected (\texttt{nfd}), a point-source flag (\texttt{PointS}), basic shape parameters (\texttt{a}, \texttt{b} \& \texttt{theta}), the signal subtracted as Background (\texttt{Backg}), Kron apertures (\texttt{rk}), fraction-of-light radii (\texttt{rf}), Petrosian apertures (\texttt{rp}), celestial coordinates of cluster's BCG (\texttt{BCG\_pos\_RA, BCG\_pos\_Dec}) and a projected physical distance (\texttt{PhyDistBCG}) from each detection to the cluster BCG in units of Mpc.  

\vspace{0.2cm}
 
 The catalogue contains a double photometry (as explained in Section \ref{RestrictedApertures}) where magnitudes (\& uncertainties) are named according to the effective filter wavelength, the camera and the adopted photometric aperture. Here we present a few examples: \texttt{F225W\_WFC3\_PHOTOZ} corresponds to the AB magnitudes for the \texttt{F225W} filter onboard the \texttt{WFC3} camera. \texttt{PHOTOZ} refers to the total \texttt{restricted} apertures used to derive photo-z estimations (Section \ref{RestrictedApertures}). Likewise, \texttt{F225W\_WFC3\_MASS} corresponds to the AB magnitudes for the \texttt{F225W} filter where \texttt{MASS} corresponds to the total \texttt{moderate} apertures used to derive absolute magnitudes and stellar masses. Photometric uncertainties take the same name (as magnitudes) but adding the prefix ``\texttt{d}'': \texttt{dF225W\_WFC3\_PHOTOZ} \& \texttt{dF225W\_WFC3\_MASS}. For both sets of magnitudes, photometric uncertainties are empirically corrected (Section \ref{photoerr1} \& \ref{photoerr2}) in all the 16 bands. Whenever a source was not detected, its magnitude was set to \texttt{99.} and its photometric uncertainty replaced by a 3-${\sigma}$ upper limit (Section \ref{upperlimits}) suitable for \texttt{BPZ}. Magnitudes are corrected from galactic extinction using \cite{1998ApJ...500..525S}.

\vspace{0.2cm}

 The best photometric redshift estimate for every source is \texttt{zb}. Additionally, \texttt{zb\_min} and \texttt{zb\_max} represent the lower and upper limits for the first peak within a $1\sigma$ interval. Spectral-type classification is given by \texttt{t\_b} where its number refers to the selected template as indicated in Figure \ref{BPZ}. \texttt{Odds} gives the amount of redshift probability enclosed around the main peak and \texttt{Chi2} the reduced chi-squared from the comparison between observed and predicted fluxes according to the selected template and redshift. Likewise, the cluster redshift in every field is also included (\texttt{clusterz}). An estimation of each detection stellar mass content (in units of log10($M\odot$)) is given by \texttt{Stell\_Mass}. Absolute magnitudes in the Johnson B-band (\texttt{M\_B}) are estimated for each detection according to its most likely redshift and spectral-type from the \texttt{BPZ} code.

\vspace{0.2cm}

In order to properly estimate stellar masses and absolute magnitudes for the background galaxies, the catalogue contains an estimate of the expected magnification due to the gravitational lensing effect. These estimates (\texttt{mu\_lens}) and its uncertainties (\texttt{dmu\_lens}) are computed from the mass models derived by \cite{2015ApJ...801...44Z} using the photometric apertures defined in section $\ref{RestrictedApertures}$ and a 1-$\sigma$ interval around the most likely redshift. A secondary estimation of both the stellar mass and the absolute magnitudes (\texttt{M\_B\_LensCor} \& \texttt{Stell\_Mass\_LensCor}) is derived for all galaxies correcting for the expected magnification at that position. Finally, the catalogue includes a compilation of 1257 spectroscopic redshifts within the WFC3 FoV, indicating the exact values (\texttt{specz\_value}), the references to the observational programme (\texttt{specz\_refer}) and the corresponding quality-flags (\texttt{specz\_qual}).

\vspace{0.2cm}

This new photometric redshift catalogue along with additional value-added products (such as PSF-models or redshift \texttt{PDFs}) can be reached from the Mikulski Archive for Space Telescopes (\texttt{MAST}) archive: \textcolor{blue}{https://archive.stsci.edu/prepds/clash/}.

\section{Summary}
\label{summary}

In this paper we have thoroughly discussed the complexity of deriving accurate colors for galaxies in dense environments and presented a new photometric pipeline for the CLASH survey capable to retrieve precise photometric redshift estimates for galaxies in massive galaxy clusters. The photo-z catalogue presented in this work represents (almost) a factor of 2 improvement in the photo-z quality with respect to the results presented in J14 (or up to a factor of 5 for galaxies with a magnitude F814W$<$20). We have processed the entire CLASH dataset with this new pipeline. This dataset is composed of 16-filter HST imaging (0.2 - 1.7 $\mu$m) of 25 massive galaxy clusters of 0.1$<$z$<$0.9. 

\vspace{0.2cm}

Main differences with respect to our previous pipeline are 1. the definition of a new set of photometric apertures (\texttt{restricted} \& \texttt{moderate}) to enhance the signal-to-noise of cluster galaxies in the bluest filters, 2. a PSF homogenization of images based on empirical PSF-models, 3. an efficient subtraction of the ICL signal from images to restore the original colors of galaxies, 4. an empirical re-estimation of photometric uncertainties in images, 5. an improved definition of photometric upper-limits fixing a previously unnoticed bias in the redshift distribution ($n(z)$) of background galaxies, 6. the extension of the \texttt{BPZ} library of templates to be able to reproduce the extreme colors of several cluster galaxies and 7. a recalibration of photometric uncertainties for very bright galaxies fixing another previously unnoticed bias affecting the photo-z quality of bright galaxies with high \texttt{Odds}. 

\vspace{0.2cm}

We have computed and quantified the accuracy reached by our new photo-z pipeline. Based on a control sample of 382 cluster galaxies spread over the 25 clusters, these photo-z estimates reach a precision of dz/1+z $\sim$0.8\% for galaxies with a magnitude F814W$<$18, a dz/1+z $\sim$1.0\% for galaxies with F814W$<$20 or a dz/1+z $\sim$2.0\%  for galaxies with F814W$<$23. According to the spectroscopic sample used in this work (see Section \ref{GCphotoz}), the overall precision is dz/1+z $\sim$2.0\%, negligible bias ($\mu$=0.004) and a fraction of catastrophic outliers always below 3\% except for the faintest magnitude bin ($23.5<m<24.5$) where the signal-to-noise of galaxies makes the photo-z estimations more uncertain. Finally, it is worth recalling that these numbers have been derived using a bright ($<F814W>21.3$) sample of spectroscopic galaxies. Therefore, its extrapolation to the faintest galaxies may be done with caution due to unwilling selection functions. As emphasized in this work, these results represent an overall factor 2 improvement for the CLASH photo-z (or up to a factor of 4 for very bright galaxies) with respect to our results presented in J14. Meanwhile, this new photometry not only provides an overall higher precision at all magnitudes and redshift ranges, with respect to our previous pipeline, but also yields an expected dependence between the photo-z precision and the signal-to-noise of galaxies, not observed in our previous version. Therefore, the results presented in this work partially alleviates the previous tension between the expected performance for the CLASH photo-z reported in P12 and the results presented in J14.

\vspace{0.2cm}

As throughly discussed in Section \ref{pzp}, we have discovered a significant tension between the NIR colors predicted by the \texttt{BPZ} models and those observed in real data. For the reddest filter (F160W) and faintest galaxies, the observed magnitude offsets suggest that galaxies in the NIR are systematically (up to $\sim$0.08) fainter than expected. When this effect is accounted for the overall photo-z precision is dramatically improved.   

\vspace{0.2cm}   

We have run \texttt{BPZ2.0} on this new CLASH photometry to generate a new photometric redshift catalogue for the 25 CLASH massive galaxy clusters. The catalogue includes astrometric, morphologic, photometric and photometric-redshift information along with an estimate of the demagnified stellar mass (\texttt{M\_B\_LensCor}) and absolute magnitudes (\texttt{Stell\_Mass\_LensCor}) for all detected sources in a NIR detection image (e.g., a weighted sum of WFC3/IR images: F105W, F110W, F125W, F140W, F160W). Based on this improved photometry and photo-z estimations, accurate redshift probability distribution functions (\texttt{PDFs}) are derived and used to carry out membership analysis in all the 25 clusters. The analysis, presented in a separate paper (Molino et al., in prep), finds $>$3500 cluster member candidates down to a magnitude M$_{B}<-13$. This sample enables the characterization of the physical properties of the faintest cluster galaxies in the CLASH fields. 

\vspace{0.2cm}

Finally, it is worth recalling that photo-z codes are very sensitive tools to the quality of the input photometry. In the era of large photometric redshift surveys a special effort must be taken in the derivation of accurate and homogeneous photometric datasets. The analysis techniques developed here will be useful in other surveys of crowded fields, including the Frontier Fields, surveys carried out with JWST, DES (\citealt{2016MNRAS.460.1270D}), Euclid (\citealt{2010arXiv1001.0061R}) or for the new generation of multi-narrow band photometric redshift surveys such as, for example, the Javalambre-Physics of the Accelerated Universe Astrophysical Survey (J-PAS; \citealt{2014arXiv1403.5237}) aiming at measuring extremely precise (dz/1+z$\sim$0.3\%) photometric redshifts on hundreds of thousands of galaxy clusters and groups \citep{2016MNRAS.456.4291A}.  

\section*{Acknowledgments}

We acknowledge the financial support of the Brazilian funding agency FAPESP (Post-doc fellowship - process number 2014/11806-9). Likewise, we acknowledge the Spanish Ministerio de Educaci\'on y Ciencia through grant AYA2006-14056  BES-2007-16280. We acknowledges the financial support from the Spanish MICINN under the Consolider-Ingenio 2010 Programme grant CSD2006-00070: First Science with the GTC. AM acknowledges the CEFCA (Centro de Estudios de F\'isica del Cosmos de Arag\'on) and the IAA-CSIC (Instituto de Astrof\'isica de Andalucia) institutes for hosting him during the period 2015/2016. PR, MM, AMe and MN acknowledge the financial support from PRIN-INAF 2014 1.05.01.94.02. Likewise, AM acknowledges the STScI (Space Telescope Science Institute) for hosting him during August - September 2010, June - August 2011 and July 2016. We acknowledge the financial support of ESO since most of the spectroscopic redshifts used in the work were based on ESO VLT programme ID 186.A-0798. BA has received funding from the European Union's Horizon 2020 research and innovation programme under the Marie Sklodowska-Curie grant agreement No 656354. This research has made use of the NASA/IPAC Extragalactic Database (NED) which is operated by the Jet Propulsion Laboratory, California Institute of Technology, under contract with the National Aeronautics and Space Administration. We dedicate this work to the memory of Dr. Javier Gorosabel, a young Spanish astronomer who unfortunately passed away during the time this paper was written.

\vspace{1.0cm}

\noindent$^1$Instituto de Astronomia, Geof\'isica e Ci\^encias Atmosf\'ericas, Universidade de S\~ao Paulo, 05508-090, S\~ao Paulo, Brazil. \\
$^2$Instituto de Astrof\'isica de Andaluc\'ia (IAA-CSIC). Glorieta de la astronom\'ia S/N. E-18008, Granada. Spain. \\
$^3$APC, AstroParticule et Cosmologie, Universit\'e Paris Diderot, CNRS/IN2P3, CEA/lrfu, Observatoire de Paris, Sorbonne Paris Cit\'e, 10, rue Alice Domon et L\'eonie Duquet, 75205 Paris Cedex 13, France \\
$^4$Space Telescope Science Institute, 3700 San Martin Drive Baltimore, MD 21218. \\
$^5$Department of Physics and Astronomy, University College of London, Gower Street, London WC1E 6BT, UK. \\
$^6$Dark Cosmology Centre, Niels Bohr Institute, University of Copenhagen, Juliane Maries Vej 30, DK-2100 Copenhagen, Denmark. \\
$^7$University Observatory Munich, Scheinerstrasse 1, D-81679 Munich, Germany. \\
$^8$Max Planck Institute for Extraterrestrial Physics, Giessenbachstrasse, D-85748 Garching, Germany. \\
$^{9}$Department of Astrophysical Sciences, 4 Ivy Lane, Princeton, NJ 08544, USA. \\
$^{10}$Dipartimento di Fisica e Scienze della Terra, Universitˆ degli Studi di Ferrara, Via Saragat 1, I-44122 Ferrara, Italy. \\
$^{11}$Observat\'orio Nacional, COAA, Rua General Jos\'e Cristino 77, 20921-400, Rio de Janeiro, Brazil. \\
$^{12}$Department of Theoretical Physics, University of the Basque Country UPV/EHU, E-48080 Bilbao. Spain. \\
$^{13}$IKERBASQUE, Basque Foundation for Science, Alameda Urquijo, 36-5 48008 Bilbao, Spain. \\
$^{14}$Universit\"at Heidelberg, Zentrum f\"ur Astronomie, Institut f\"ur Theoretische Astrophysik, Philosophenweg 12, 69120 Heidelberg, Germany.\\
$^{15}$Leiden Observatory, Leiden University, NL-2300 RA Leiden, Netherlands.\\
$^{16}$Department of Physics, Math, and Astronomy, California Institute of Technology, Pasadena, CA 91125, USA \\
$^{17}$Physics and Astronomy Department, Michigan State University, 567 Wilson Rd., East Lansing, MI 48824, USA\\
$^{18}$Department of Physics and Astronomy, The Johns Hopkins University, 3400 North Charles Street, Baltimore, MD 21218, USA \\
$^{19}$Center for Cosmology and Particle Physics, New York University, New York, NY 10003, USA. \\
$^{20}$Department of Astronomy, University of California, Berkeley, CA 94720, USA. \\
$^{21}$Dark Cosmology Centre, Niels Bohr Institute, University of Copenhagen, Juliane Maries Vej 30, DK-2100 Copenhagen, Denmark. \\
$^{22}$Instituto de Astrof\'isica, Facultad de F\'isica, Centro de Astroingenier\'ia, Pontificia Universidad Cat\'olica de Chile Av. Vicu\~na Mackenna 4860, 782-0436 Macul, Santiago, Chile. \\ 
$^{23}$Department of Physics and Astronomy, Rutgers, The State University of New Jersey, Piscataway, NJ 08854, USA. \\
$^{24}$The Observatories, The Carnegie Institution for Science, 813 Santa Barbara St., Pasadena, CA 91101, USA. \\
$^{25}$School of Physics and Astronomy, Tel-Aviv University, Tel-Aviv 69978, Israel. \\
$^{26}$ INAF. Osservatorio Astronomico di Capodimonte, Via Moiariello 16, I-80131 Napoli, Italy
$^{27}$INAF - Osservatorio Astronomico di Bologna, via Ranzani 1, 40127 Bologna, Italy. \\
$^{28}$Department of Physics, University of Oxford, Oxford OX1 3RH, UK. \\
$^{29}$NASA Jet Propulsion Laboratory, California Institute of Technology, Pasadena, California 91109, USA \\
$^{30}$INAF - Osservatorio Astronomico di Trieste, via G. B. Tiepolo 11, I-34143, Trieste, Italy \\
$^{31}$Department of Physics and Astronomy, University of South Carolina, 712 Main St., Columbia, SC 29208, USA. \\
$^{32}$Division of Physics, Math, and Astronomy, California Institute of Technology, Pasadena, CA 91125. \\
$^{33}$Institute of Astronomy and Astrophysics, Academia Sinica, P.O. Box 23-141, Taipei 10617, Taiwan. \\
$^{34}$Physics Department, Ben-Gurion University of the Negev, P.O. Box 653, BeÕer-Sheva 84105, Israel \\
\vspace{1.0cm}

\appendix

\begin{table*}
\caption{The CLASH photometric redshift catalogs.}
\begin{center}
\label{catalogdesc1}
\begin{tabular}{|l|c|c|c|c|c|c|c|}
\hline
\hline
& COLUMN  &   PARAMETER  &  DESCRIPTION \\
\hline
& 1        		  &      CLASH\_ID	                  &  CLASH Object ID Number	 \\
& 2                      &      ClusterName	                   & CLASH Cluster Name \\
& 3                      &      RA	                                    &  Right Ascension in decimal degrees [J2000]   \\
& 4        		  &      DEC	                           &  Declination in decimal degrees [J2000]   \\
& 5                      &      XX 	                                    &  x-pixel coordinate	 \\
& 6                      &      YY 	                                    &  y-pixel coordinate	 \\
& 7                      &      AREA	                           &  Isophotal aperture area (pixels)	 \\
& 8                     &      FWHM	                           &  Full width at half maximum (arcsec)	 \\
& 9                     &      s2n                  		          &   Signal-to-Noise on (NIR) detection image \\
& 10                     &      PhotoFlag                         &   SExtractor Photometric Flag \\
& 11                   &       nfo                                     &     Number Filters Observed (out of 16)  \\
& 12                   &       nfd                                     &     Number Filters Detected (out of 16)  \\
& 13                    &      PointS                              &    Potential Point-Source [0: extended, 1: candidate] \\
& 14			  &      THETA                             &   Position Angle (CCW/x)  \\
& 15                    &      a                                      &   Profile RMS along major axis (pixels) \\
& 16                    &      b                                       &  Profile RMS along minor axis (pixels) \\ 
& 17                    &      Backg                               &   Background-signal subtracted from detections \\
& 18                    &      RK                                   &   Kron apertures in units of A or B (pixels) \\ 
& 19                    &      RF                                   &    Fraction-of-light radii (pixels)  \\
& 20                    &      RP                                   &    Petrosian radii (pixels)  \\
& 21                    &      BCG\_pos\_RA                  &  Right Ascension for the BCG in decimal degrees [J2000] \\
& 22                    &      BCG\_pos\_Dec                 & Declination for the BCG in decimal degrees [J2000] \\
& 23                    &      PhyDistBCG                    &    Projected Physical Distance to BCG (Mpc)  \\
& 24                    & F225W\_WFC3\_PHOTOZ     &  F225W/WFC3 ``restricted'' magnitude (AB); best for photo-z \\
& 25                    & dF225W\_WFC3\_PHOTOZ   &  F225W/WFC3 ``restricted'' magnitude uncertainty (AB); best for photo-z \\
& 26                    & F225W\_WFC3\_MASS          &  F225W/WFC3 ``moderated'' magnitude (AB); best for stellar mass \\
& 27                    & dF225W\_WFC3\_MASS        &  F225W/WFC3 ``moderated'' magnitude uncertainty (AB); best for stellar mass \\
&   ....                  &              ............                     &         ...............           \\ 
& 40                    & F435W\_ACS\_PHOTOZ     &  F435W/ACS ``restricted'' magnitude (AB) best for photo-z \\
& 41                    & dF435W\_ACS\_PHOTOZ   &  F435W/ACS ``restricted'' magnitude uncertainty (AB) best for photo-z \\
& 42                    & F435W\_ACS\_MASS          &  F435W/ACS ``moderated'' magnitude (AB) best for stellar mass \\
& 43                    & dF435W\_ACS\_MASS        &  F435W/ACS ``moderated'' magnitude uncertainty (AB) best for stellar mass \\
&   ....                  &              ............                     &         ...............           \\ 
& 68                    & F105W\_WFC3\_PHOTOZ     &  F105W/WFC3 ``restricted'' magnitude (AB) best for photo-z \\
& 69                    & dF105W\_WFC3\_PHOTOZ   &  F105W/WFC3 ``restricted'' magnitude uncertainty (AB) best for photo-z \\
& 70                    & F105W\_WFC3\_MASS          &  F105W/WFC3 ``moderated'' magnitude (AB) best for stellar mass \\
& 71                    & dF105W\_WFC3\_MASS        &  F105W/WFC3 ``moderated'' magnitude uncertainty (AB) best for stellar mass \\
&   ....                  &              ............                     &         ...............           \\ 
& 88                   &     clusterz                              & Cluster Redshift  \\
&  89                   &    zb                                     & \texttt{BPZ} most likely redshift  \\
&  90                  &     zb\_min                            &  Lower limit (95p confidence) \\
&  91                  &     zb\_max                            & Upper limit (95p confidence) \\
&  92                  &     tb                                     &  \texttt{BPZ} most likely spectral type \\
&  93                  &     Odds                                & P(z) contained within zb +/- 2*0.03*(1+z) \\
&  94                  &     Chi2                                 &  Poorness of \texttt{BPZ} fit: observed vs. model fluxes \\ 
&  95                  &     M\_B                             &    Absolute Magnitude in the B\_Johnson band [AB] \\
&  96                  &     Stell\_Mass                      &     Stellar Mass (log10($M_{\odot}$))  \\
&  97                  &    F814W\_ACS\_MASS\_LensCor  & Lensing-corrected F814W\_ACS\_MASS magnitude \\
&  98                  &    MB\_LensCor          &    Lensing-corrected Absolute Magnitude [AB] (B\_Johnson) \\
&  99                &     Stell\_Mass\_LensCor    &  Lensing-corrected Stellar Mass (log10($M_{\odot}$))  \\
&  100                &     specz\_value                    &  Spectroscopic Redshift [-99 if unknown]  \\
&  101              &     specz\_refer                &  Reference for the Spectroscopic Redshift  \\
&  102                &     specz\_qual                &  Quality of the Spectroscopic Redshift [0:secure, 1:likely, 2: unsure] \\
\hline
\hline
\end{tabular}
\end{center}
\end{table*}
%&  95                  &     mu\_lens                        &    Averaged Flux Magnification at detection position \\
%&  96                  &     dmu\_lens                     &    Flux Magnification Uncertainty at detection position \\

\vspace{0.5cm}

\begin{table}
\caption{CLASH photometric filter system. The table includes the filter effective wavelengths ($\lambda_{eff}$), the filter widths (FWHM), the averaged limiting magnitudes (m$_{lim}$) measured in 3" circular apertures and a 5-$\sigma$ significance and the averaged exposure time per each filter.}
\begin{center}
\label{CLASHFS}
\begin{tabular}{|l|c|c|c|c|c|c|c|clclclclcl}
\hline
\hline
FILTER/    &  $\lambda_{eff}$ &  FWHM     &  $\langle$m$_{lim}^{(3")}$$\rangle$ & $<$Time$>$  \\
CAMERA  &            [$\AA$]     &  [$\AA$]   &   (5-$\sigma$)  \\ 
\hline
UltraViolet \\
\hline
F225W/WFC3  &  2414.5  &  67.7 &  26.4 & 3558  \\
F275W/WFC3  &  2750.3  & 88.7  &  26.5 & 3653  \\
F336W/WFC3  & 3381.2  & 203.6 &  26.6 & 2348  \\
F390W/WFC3  & 3956.4 & 526.9  &  27.2 & 2350 \\
\hline
Optical \\
\hline
F435W/ACS  &  4365.2 & 203.3 &  27.2 & 1984  \\
F475W/ACS  &  4842.5 & 318.7 &  27.6 & 1994 \\
F606W/ACS  &  6104.6 & 809.9 &  27.6 &  1975  \\
F625W/ACS  &  6374.6 & 546.0 &  27.2 & 2022  \\
F775W/ACS  &  7755.2 & 603.0 &  27.0 & 2022   \\
F814W/ACS  &  8454.5 & 2366.4 &  27.7& 4103  \\
F850LP/ACS  &  9140.7 & 388.4 &  27.7 & 4045  \\
\hline
Near-Infrarred \\
\hline
F105W/WFC3  &  11923.1 & 2818.5 &  27.3 & 2645  \\
F110W/WFC3  &  12249.8 & 4594.5 &  27.8 & 2415  \\
F125W/WFC3 &  12610.5 & 2844.4 &  27.2 & 2425  \\
F140W/WFC3  &  14153.4 & 4010.0 &  27.4 & 2342  \\
F160W/WFC3  &  15523.4 & 2855.5 &  27.5 & 4920  \\
\hline
\hline
\end{tabular}
\end{center}
\end{table} 

\begin{table*}
\caption{Photometric zero-point offsets (I). The table includes the observed magnitude offsets and errors as much for three different families of objects (foreground, cluster and background galaxies) as for the total (Global) sample. Values are normalized to the ACS/F775W filter. The number of galaxies utilized in every bin ($\#$) are indicated in the bottom part of the table.}
\begin{center}
\label{CLASHFS1}
\begin{tabular}{|l|c|c|c|c|c|c|c|clclclclcl}
\hline
\hline
FILTER  & Foreground & Cluster & Background & Global \\
\hline
\hline
F435W  & -0.05 $\pm$ 0.09 & -0.15 $\pm$ 0.14 & -0.16 $\pm$ 0.14 & -0.14 $\pm$ 0.13 \\
F475W  & -0.08 $\pm$ 0.05 & -0.10 $\pm$ 0.08 & -0.12 $\pm$ 0.09 & -0.11 $\pm$ 0.08 \\
F606W  & -0.01 $\pm$ 0.04 &  0.02 $\pm$ 0.05 & -0.03 $\pm$ 0.07 & -0.01 $\pm$ 0.06 \\
F625W  & -0.01 $\pm$ 0.06 &  0.01 $\pm$ 0.05 & -0.01 $\pm$ 0.10 &  0.00 $\pm$ 0.07 \\
F775W  &  0.00 $\pm$ 0.04 &  0.00 $\pm$ 0.04 &   0.00 $\pm$ 0.08 &  0.00 $\pm$ 0.06 \\
F814W  & -0.01 $\pm$ 0.03 &  0.00 $\pm$ 0.04 & -0.02 $\pm$ 0.07 & -0.01 $\pm$ 0.05 \\
F850LP &  0.01 $\pm$ 0.04 & -0.01 $\pm$ 0.03 & -0.01 $\pm$ 0.08 & -0.02 $\pm$ 0.05 \\
F105W  & -0.02 $\pm$ 0.02 & -0.03 $\pm$ 0.02 & -0.02 $\pm$ 0.05 & -0.03 $\pm$ 0.03 \\
F110W  & -0.03 $\pm$ 0.03 & -0.04 $\pm$ 0.02 & -0.04 $\pm$ 0.04 & -0.04 $\pm$ 0.03 \\
F125W  & -0.07 $\pm$ 0.02 & -0.07 $\pm$ 0.01 & -0.08 $\pm$ 0.03 & -0.08 $\pm$ 0.02 \\
F140W  & -0.07 $\pm$ 0.01 & -0.07 $\pm$ 0.01 & -0.10 $\pm$ 0.02 & -0.09 $\pm$ 0.01 \\
F160W  & -0.06 $\pm$ 0.02 & -0.06 $\pm$ 0.02 & -0.12 $\pm$ 0.03 & -0.09 $\pm$ 0.02 \\
\hline
Sample &              99           &              561           &            635           &           1295          \\
\hline
\hline
\end{tabular}
\end{center}
\end{table*}

\begin{table*}
\caption{Photometric zero-point offsets (II). Same as in the previous table but dividing the global sample in three different magnitude (F814W) bins.}
\begin{center}
\label{CLASHFS2}
\begin{tabular}{|l|c|c|c|c|c|c|c|clclclclcl}
\hline
\hline
FILTER  & $17<m_{F814W}<21$ & $21<m_{F814W}<23$ & $23<m_{F814W}<25$ & Global \\
\hline
\hline
F435W  &  -0.08 $\pm$ 0.07 & -0.09 $\pm$ 0.07 & -0.12 $\pm$ 0.06  & -0.14 $\pm$ 0.13 \\
F475W  & -0.06 $\pm$ 0.04 & -0.07 $\pm$ 0.07 & -0.11 $\pm$ 0.06 & -0.11 $\pm$ 0.08 \\
F606W  & -0.01 $\pm$ 0.03 & -0.02 $\pm$ 0.05 & -0.02 $\pm$ 0.04 & -0.01 $\pm$ 0.06 \\
F625W  &  0.00 $\pm$ 0.03 & -0.00 $\pm$ 0.05 &  -0.01 $\pm$ 0.06 & -0.00 $\pm$ 0.07 \\
F775W  &  0.00 $\pm$ 0.02  & 0.00 $\pm$ 0.04  &  0.00 $\pm$ 0.05 & 0.00 $\pm$ 0.06 \\
F814W  &  0.01 $\pm$ 0.02  & -0.01 $\pm$ 0.03 &  -0.02 $\pm$ 0.04 & -0.01 $\pm$ 0.05 \\
F850LP & 0.01 $\pm$ 0.02  & -0.02 $\pm$ 0.03 &  -0.02 $\pm$ 0.05 & -0.02 $\pm$ 0.05 \\
F105W  & -0.01 $\pm$ 0.02 & -0.03 $\pm$ 0.02 & -0.02 $\pm$ 0.04 & -0.03 $\pm$ 0.03 \\
F110W  & -0.02 $\pm$ 0.01 & -0.05 $\pm$ 0.02 & -0.05 $\pm$ 0.03 & -0.04 $\pm$ 0.03 \\
F125W  & -0.04 $\pm$ 0.01 & -0.08 $\pm$ 0.02 & -0.10 $\pm$ 0.03 & -0.08 $\pm$ 0.02 \\
F140W  & -0.03 $\pm$ 0.02 & -0.09 $\pm$ 0.02 &  -0.12 $\pm$ 0.03 & -0.09 $\pm$ 0.01 \\
F160W  & 0.00 $\pm$ 0.03 & -0.08 $\pm$ 0.03 &  -0.14 $\pm$ 0.05 & -0.09 $\pm$ 0.02 \\
\hline
Sample &            207            &            337           &            351            &           1295          \\
\hline
\hline
\end{tabular}
\end{center}
\end{table*}   

\begin{table*}
\caption{Photometric zero-point offsets (III). Same as in previous tables but dividing the global sample in three different redshift bins.}
\begin{center}
\label{CLASHFS3}
\begin{tabular}{|l|c|c|c|c|c|c|c|clclclclcl}
\hline
\hline
FILTER  & $z_{s}<0.3$ & $0.3<z_{s}<0.5$ &  $z_{s}>0.5$ & Global \\
\hline
\hline
F435W  & -0.04 $\pm$ 0.04 & -0.07 $\pm$ 0.06 & -0.09 $\pm$ 0.05 & -0.14 $\pm$ 0.13 \\
F475W  & -0.04 $\pm$ 0.03 & -0.07 $\pm$ 0.06 & -0.08 $\pm$ 0.07 & -0.11 $\pm$ 0.08 \\
F606W  &  0.02 $\pm$ 0.02 &  0.01 $\pm$ 0.03 & -0.02 $\pm$ 0.04 & -0.01 $\pm$ 0.06 \\
F625W  &  0.01 $\pm$ 0.02 &  0.01 $\pm$ 0.03 &  0.02 $\pm$ 0.06  &  0.00 $\pm$ 0.07 \\
F775W  &  0.00 $\pm$ 0.02 &  0.00 $\pm$ 0.03 &  0.00 $\pm$ 0.04  &  0.00 $\pm$ 0.06 \\
F814W  &  0.01 $\pm$ 0.02 & -0.00 $\pm$ 0.02 &  0.00 $\pm$ 0.04 & -0.01 $\pm$ 0.05 \\
F850LP &  0.01 $\pm$ 0.02 & -0.01 $\pm$ 0.02 & -0.01 $\pm$ 0.03 & -0.02 $\pm$ 0.05 \\
F105W  & -0.00 $\pm$ 0.02 & -0.02 $\pm$ 0.02 & -0.02 $\pm$ 0.03 & -0.03 $\pm$ 0.03 \\
F110W  & -0.01 $\pm$ 0.02 & -0.04 $\pm$ 0.02 & -0.04 $\pm$ 0.03 & -0.04 $\pm$ 0.03 \\
F125W  & -0.03 $\pm$ 0.01 & -0.06 $\pm$ 0.02 & -0.08 $\pm$ 0.02 & -0.08 $\pm$ 0.02 \\
F140W  & -0.00 $\pm$ 0.03 & -0.07 $\pm$ 0.02 & -0.10 $\pm$ 0.03 & -0.09 $\pm$ 0.01 \\
F160W  & -0.00 $\pm$ 0.03 & -0.04 $\pm$ 0.03 & -0.10 $\pm$ 0.03 & -0.09 $\pm$ 0.02 \\
\hline
\,\,\,$\#$  &            91             &            945           &             259           &          1295          \\
\hline
\hline
\end{tabular}
\end{center}
\end{table*} 

\label{lastpage}
\end{document}